\documentclass[onecolumn]{article}

\usepackage{enumitem}
\usepackage{amsmath}
\usepackage{bm}
\usepackage{graphicx}
\usepackage{color}
\usepackage{nomencl}
\usepackage{xstring}
\usepackage{tabto}
\usepackage{authblk}

\newcommand{\nomenclheader}[1]{
\item[\hspace*{-\itemindent}\normalfont\bfseries #1]}
\renewcommand\nomgroup[1]{
        \IfStrEqCase{#1}{
	{A}{\nomenclheader{Acronyms}}
	{D}{\nomenclheader{Dimensional Properties}}
	{N}{\nomenclheader{Non-dimensional Properties}}
	{O}{\nomenclheader{Operators}}
	{S}{\nomenclheader{Subscripts and Superscripts}}
}
}

\makenomenclature 

\begin{document}

\title{Numerical Evaluation of  Entropy Generation \\ in Isolated Airfoils and Wells Turbines}

\author[1]{Tiziano Ghisu}
\author[1]{Francesco Cambuli}
\author[1]{Pierpaolo Puddu} 
\author[1]{Natalino Mandas}
\author[2]{Pranay Seshadri}
\author[2]{Geoffrey T. Parks}
\affil[1]{Department of Mechanical, Chemical and Materials Engineering, University of Cagliari, via Marengo 2, 09123 Cagliari, Italy}
\affil[2]{Department of Engineering, University of Cambridge, Trunmpington Street, Cambridge CB21PZ, United Kingdom}
\setcounter{Maxaffil}{0}
\renewcommand\Affilfont{\itshape\small}



\maketitle

\begin{abstract}
In recent years, a number of authors have studied entropy generation in Wells turbines. This is potentially a very interesting topic, as it can provide important insights into the irreversibilities of the system, as well as a methodology for identifying, and possibly minimizing, the main sources of loss. Unfortunately, the approach used in these studies contains some crude simplifications that lead to a severe underestimation of entropy generation and, more importantly, to misleading conclusions. 

This paper contains a re-examination of the mechanisms for entropy generation in fluid flow, with a particular emphasis on RANS equations. 
An appropriate methodology for estimating entropy generation in isolated airfoils and Wells turbines is presented.
Results are verified for different flow conditions, and a comparison with theoretical values is presented.
\end{abstract}

\section{Introduction}

Sea-wave energy has the highest availability among renewable energy sources, and could provide an important contribution to meeting the demand for electricity worldwide, while reducing the dependency on fossil fuels \cite{Pelc2002}. An effective method for converting sea-wave energy into electrical energy consists of the combination of a chamber that produces an alternating motion of a mass of air (an Oscillating Water Column, or OWC, system~\cite{Boccotti2007,Boccotti2012}), and a turbine that converts this energy into shaft power \cite{Thorpe2000,Taylor1983,Falcao2010}). 
Typically, the latter is achieved with a Wells turbine ~\cite{Wells,Raghunathan1995,Gato1988}, defined by a simple but reliable layout: a rotor with symmetric blades staggered at 90 degrees with respect to the incoming flow. This configuration provides equal performance in both outflow (air is flowing out of the chamber), and inflow (air flowing into the chamber).

Wells turbines have been studied extensively, both experimentally \cite{Curran1997,Govardhan1998,Gato2001,Setoguchi2003a,Thakker2007a,Paderi2013a,Puddu2014,Moisel2016} and numerically~\cite{Kim2001,Setoguchi2003b,Dhanasekaran2005,Torresi2009,Gomes2012,Starzmann2013,Halder2015,Ghisu_JFE}, but mainly from a first-law perspective. More recently, a number of authors~\cite{Shehata2016,Shehata2017,Shehata2017a,Shehata2017b,Soltanmohamadi2016} have focused on the irreversibilities in this component, trying to link the amount of available work (exergy) and the entropy produced. 

The application of second-law analyses is not novel, as such analyses have been done by many authors, but mainly from a system perspective: Bejan~\cite{Bejan1987,Bejan1996} presents an extensive review of the application to different thermodynamic systems. Similar approaches have been used in the analysis of thermal power plants~\cite{Kaushik2011}, wind turbines \cite{Pope2010,Baskut2010,Baskut2011,Redha2011,Ozgener2007}, and vortex generators~\cite{Herpe2009}.

Computational Fluid Dynamics (CFD), aside from making possible a more rapid and economic evaluation of many systems (compared to experimental testing), can provide a better and more detailed understanding of many phenomena, by capturing a level of detail that would be extremely difficult in an experimental study. It allows a very fine decomposition of the overall problem that can be used to locate the sources of irreversibility within the system. Entropy generation by thermal and viscous sources can be calculated directly by post-processing the fields of thermo- and fluid-dynamic variables available from the numerical solution~\cite{Bejan1987}. However, particular attention needs to be paid to the nature of the flow and to the solution approach. While in laminar flows all entropy generation is directly found in the CFD solution~\cite{Drost1991,Abuhikleh1999,Shuja2001,Herpe2009,Wahba2011,Rashidi2013}, this is not true for turbulent flows, unless a Direct Numerical Simulation (or DNS) is used. However, with current computational resources, the use of DNS is still impractical even for moderately large Reynolds numbers. To make high-Reynolds-number flows treatable, the governing equations are either averaged (leading to the Reynolds Averaged Navier Stokes equations, or RANS) or filtered (leading to the Large Eddy Simulation approach, or LES).

Moore and Moore~\cite{Moore1983,Moore1983a} present a methodology for calculating entropy production in viscous flows, using RANS equations. They divide the mean entropy production into the contributions due to mean flow and turbulent fluctuations: as only the former is directly resolved by the numerical method, the latter has to be modeled, and in \cite{Moore1983,Moore1983a} this is done using  a mixing-length turbulence model. In~\cite{Moore1983} this approach is validated by comparison with experimental data from turbulent boundary layers, while in~\cite{Moore1983a} the methodology is used to calculate and isolate the source of entropy generation in a three-dimensional channel with 90 degrees of turning.
Kock and Herwig~\cite{Kock2004} investigate entropy production in RANS equations, dividing it again into the two contributions due to mean flow and turbulent fluctuations. Since turbulent fluctuations are not directly solved in RANS, their contribution needs to be modeled, as in the other conservation equations, i.e. momentum and energy. Results from a turbulent pipe flow are presented and compared to DNS solutions, demonstrating how, even at a moderate Reynolds number of $\sim$14,000, the contribution of the turbulent entropy production is considerable and can be estimated from RANS models. Kis and Herwig~\cite{Kis2011} present results from DNS simulation of forced convection in channel flows (Re$\simeq$59,000), showing how the contribution of mean entropy production is negligible when compared to turbulent entropy production. 
The same approach of~\cite{Kock2004} is used in~\cite{Ertesvag2005}, where it is extended to turbulent combustion flows.
Adeyinka and Naterer~\cite{Adeyinka2005} and Iandoli \emph{et al.}~\cite{Iandoli2008} use the Boussinesq's hypothesis to estimate the entropy production in two- and three-dimensional turbomachinery flows: local and global entropy production values are compared with overall isentropic efficiencies, with reasonable agreement. The validity of this assumption is discussed in detail by Asinari~\emph{et al.} \cite{Asinari2016}, with particular emphasis on the $k-\epsilon$ model. 

All these studies pointed out the importance of turbulent entropy production: even at moderate Reynolds numbers ($\sim$10$^4$), the entropy produced by turbulent fluctuation is considerable, and neglecting it would lead to an underestimation of losses. Nevertheless, in a large body of literature, entropy generation in the presence of turbulent flows is calculated considering only the contribution of the mean flow. This simplification has been adopted for Wells turbines \cite{Shehata2016,Shehata2017,Shehata2017a,Shehata2017b,Soltanmohamadi2016}, pipe flows~\cite{Saqr2016,Lakzian2017}, and turbine blades~\cite{Natalini1999}. Herwig \emph{et al.}~\cite{Herwig2006} demonstrated how neglecting the entropy produced by turbulent fluctuation may lead to errors in the global entropy production, as well as to misleading conclusions. 

In light of the growing interest in second-law methods in turbomachinery analysis and optimisation, and of the questionable assumptions of several recent works, this paper presents a re-examination of entropy generation in Wells turbines. Particular attention is dedicated to the mechanisms of entropy generation and to its numerical modeling through the solution of RANS equations. A methodology for entropy quantification is presented and applied to two test cases~\cite{Shehata2016,Soltanmohamadi2016}, already studied in the past by other authors, with assumptions and simplifications that led to incorrect and physically implausible results.

This manuscript is organized as follows: Section \ref{sec:entropy} presents a review of entropy generation analysis with the Navier-Stokes equations. Section~\ref{sec:entropy2} presents an analysis of the effect of Reynolds averaging on Navier-Stokes equations and entropy production. Section~\ref{sec:entropy_airfoils} introduces an intermediate test case, i.e. the flow around an isolated airfoil, which provides an ideal problem to verify the theoretical analysis. Section~\ref{sec:methods} presents the numerical approach, Section~\ref{sec:results} presents the findings of this work. Finally, Section~\ref{sec:conclusion} draws some conclusions.

\section{Entropy Generation in Fluid Flow}
\label{sec:entropy}
In any point of a convective field, irreversibilities are produced by two distinct mechanisms: heat flow and friction. Bejan~\cite{Bejan1987}, Adeyinka and Naterer~\cite{Adeyinka2005} and Asinari~\emph{et al.} \cite{Asinari2016} give comprehensive reviews of the theory. A brief summary is presented here.

The laws of conservation of mass, momentum and energy, in differential form, can be expressed as follows:
\begin{eqnarray}
{\color{white}\frac{\partial \rho}{\partial \rho}+}\nabla \cdot {\bf{u}} &=& 0 
\label{eq:NS_ma} \\
\rho \frac{\partial {\bf{u}}}{\partial t} + \rho \nabla \cdot \left({\bf{u}} \otimes {\bf{u}}\right) = \rho \frac{D {\bf{u}}}{Dt} &=& -\nabla p + \nabla \cdot {\bf{\Pi}} + \rho{\bf{a}}
\label{eq:NS_mo} \\
\rho \frac{\partial h}{\partial t} + \rho \nabla \cdot \left({\bf{u}} h\right) = \rho \frac{D h}{Dt} &=& \frac{Dp}{Dt}-\nabla \cdot {\bf{q}} + {\bf{\Pi}} \colon \nabla{\bf{u}}
\label{eq:NS_en}
\end{eqnarray}
where $\rho$ is the fluid density (assumed constant), ${\bf{u}}$ is the fluid velocity, $p$ is the pressure, ${\bf{\Pi}}$ is the viscous part of the stress tensor, $h$ is the enthalpy, ${\bf{q}}$ is the thermal flux, and ${\bf{a}}$ is an external acceleration (e.g. gravity). The previous system of equations is not closed, and some empirical expressions are required for ${\bf{q}}$ and ${\bf{\Pi}}$. Assuming the thermal flux to obey Fourier's law and the flow to be Newtonian:
\begin{eqnarray}
{\bf{q}}&=&-\lambda \nabla T 
\label{eq:NS_fou} \\
{\bf{\Pi}}&=& \mu \left( \nabla {\bf{u}}+ \nabla {\bf{u}}^T \right) = 2 \mu \nabla^S {\bf{u}}
\label{eq:NS_bou}
\end{eqnarray}
where $\lambda$ is the thermal conductivity, $\mu$ the dynamic viscosity (both assumed constant) and $T$ the flow temperature. The system of equations can now be closed with two equations of state linking $p$, $T$, and $h$.

Entropy can be related to other thermodynamic properties by Gibbs' relation:
\begin{equation}
T \frac{Ds}{Dt}=\frac{Dh}{Dt}-\frac{1}{\rho}\frac{Dp}{Dt}
\label{eq:gibbs}
\end{equation}
Substituting equations ({\ref{eq:NS_en}, \ref{eq:NS_fou}, \ref{eq:NS_bou}) into equation (\ref{eq:gibbs}):
\begin{equation}
T \frac{Ds}{Dt}=\frac{\lambda}{\rho}\nabla^2{T}+2\frac{\mu}{\rho} \left(\nabla^S {\bf{u}} \right)^2
\end{equation}

The same can be also written as:
\begin{equation}
\frac{Ds}{Dt}+\nabla \cdot \left(-\frac{\lambda}{\rho T} \nabla T\right) = \frac{\lambda}{\rho T^2}\left(\nabla T\right)^2 + \frac{2\mu}{\rho T}\left(\nabla^S {\bf{u}}\right)^2
\end{equation}
or equivalently:
\begin{equation}
\frac{\partial s}{\partial t}+\nabla \cdot \left(s{\bf{u}}-\frac{\lambda}{\rho T} \nabla T\right) = \sigma_{T}+\sigma_{V}
\label{eq:entropy}
\end{equation}
where:
\begin{eqnarray}
\sigma_T=\frac{\lambda}{\rho T^2}\left(\nabla T\right)^2 \ge 0 
\label{eq:sigmat}\\
\sigma_V=\frac{2\mu}{\rho T}\left(\nabla^S {\bf{u}}\right)^2 \ge 0
\label{eq:sigmav}
\end{eqnarray}
are the entropy production rates per unit mass due to heat transfer and fluid flow, respectively. In accordance with the second law of thermodynamics, both terms are non-negative by definition.

Equation~(\ref{eq:NS_en}) can be also rewritten as:
\begin{eqnarray}
\rho \frac{Dh}{Dt}- \frac{Dp}{Dt}=\rho c_v \frac{D T}{Dt}&=&-\nabla \cdot {\bf{q}} + {\bf{\Pi}}\colon \nabla {\bf{u}}\nonumber \\
&=& \lambda \nabla^2 T + 2\mu(\nabla^S {\bf{u}})^2 \nonumber \\
&=& \lambda \nabla^2 T + \rho T \sigma_V
\end{eqnarray}
where it appears clear that velocity gradients act to dissipate kinetic energy into internal energy. The term $2 (\nabla^S {\bf{u}})^2$ is often called dissipation function and referred to with the greek letter $\Phi$.

\begin{eqnarray}
\Phi=2(\nabla^S {\bf{u}}) &=&
 2 \left[ \left( \frac{\partial u}{\partial x} \right)^2 +\left(\frac{\partial v}{\partial y}\right)^2+\left(\frac{\partial w}{\partial z}\right)^2\right] + \nonumber \\
&+& \left(\frac{\partial u}{\partial y}+\frac{\partial v}{\partial x} \right)^2 + \left(\frac{\partial u}{\partial z}+\frac{\partial w}{\partial x} \right)^2 + \nonumber \\
&+& \left(\frac{\partial v}{\partial z}+\frac{\partial w}{\partial y} \right)^2 
\end{eqnarray}

Similar considerations can be drawn in the case of compressible flows, as reported in~\cite{Asinari2016a}, but are outside the scope of this work.

\section{Estimating Entropy Generation by Means of Computational Fluid Dynamics}
\label{sec:entropy2}
While equations~(\ref{eq:NS_ma}, \ref{eq:NS_mo}, \ref{eq:NS_en}, \ref{eq:entropy}) are of general validity, in the presence of a turbulent flow, which is the case for the vast majority of problems of industrial interest, viscous dissipation occurs at different scales, ranging from large scales, of the same order of magnitude as the problem under investigation, to the Kolmogorov scales, which can be remarkably small, especially in the presence of high-Reynolds-number flows~\cite{Mathieu2000}. Under these conditions, the solution of the Navier-Stokes equations would require a spatial mesh and a time step sufficiently small to resolve even the smallest turbulent eddies and the fastest fluctuations (this is the approach of a DNS). This is impractical in most cases, and the governing equations are either time-averaged (leading to the RANS equations) or spatially filtered (leading to LES): in either case, not all of the turbulent scales are solved and at least a fraction (in LES, or all in RANS) of them and of their effect on the mean flow need to be modeled. The treatment of turbulence in RANS equations is described in the following section.

\subsection{Turbulence Modeling in RANS Equations}

RANS equations are based on the assumption that every variable can be expressed as the sum of a mean and a fluctuating component with zero mean value: 
\begin{equation}
\varphi(t)=\overline{\varphi (t)}+\varphi'(t)
\end{equation}
where
\begin{equation}
\overline{\varphi (t)}=\frac{1}{\tau}\int_t^{t+\tau} \varphi(t) dt
\end{equation}
and 
\begin{equation}
\overline{\varphi'(t)}=\frac{1}{\tau}\int_t^{t+\tau} \varphi'(t) dt=0
\end{equation}

Substituting in equations (\ref{eq:NS_ma}, \ref{eq:NS_mo}, \ref{eq:NS_en}), and performing a time average of the conservation equations, leads to the derivation of the RANS equations for flows of constant density and viscosity:

\begin{eqnarray}
{\color{white}\frac{\partial \rho}{\partial \rho}+} \nabla \cdot \overline{\bf{u}} &=& 0 
\label{eq:RANS_ma} \\
\rho \frac{\partial \overline{{\bf{u}}}}{\partial t} +  \rho \nabla \cdot \left(\overline{\bf{u}} \otimes {\overline{\bf{u}}}+\overline{{\bf{u}}' \otimes {\bf{u}}'}\right) &=& -\nabla \overline{p} + \nabla \cdot {\overline{\bf{\Pi}}} + \nonumber \\ 
&& + \rho \overline{{\bf{a}}} 
\label{eq:RANS_mo} \\
\rho \frac{\partial \overline{h}}{\partial t} + \rho \nabla \cdot \left({\overline{\bf{u}}} \overline{h} + \overline{{\bf{u}}' h'}\right) &=& \lambda \nabla^2 \overline{T} + \overline{{\bf{\Pi}}} \colon \nabla{\overline{\bf{u}}}+ \nonumber \\
&& + \overline{{\bf{\Pi}}' \colon \nabla{\bf{u}}'} + \frac{D\overline{p}}{Dt}
\label{eq:RANS_en}
\end{eqnarray}

The terms $\rho \nabla \cdot \left(\overline{{\bf{u}}' \otimes {\bf{u}}'}\right)$, 
$\rho \nabla \cdot \left(\overline{{\bf{u}}' h'}\right)$, 
and $\overline{{\bf{\Pi}}' \colon \nabla{\bf{u}}'}$
represent the effects of turbulent fluctuations (not resolved by the RANS approach) on the mean flow, and need somehow to be modeled. Linear eddy viscosity models are the most common approach to turbulence closure and link the above terms to the gradients of the mean variables (the so-called Boussinesq's hypothesis):

\begin{eqnarray}
-\rho \overline{{\bf{u}}' \otimes {\bf{u}}'} = {\bf{\Pi}}_R &=& \mu_t \left(\nabla {\overline{\bf{u}}} + \nabla {\overline{\bf{u}}}^T\right) = 2 \mu_t \nabla^S \overline{{\bf{u}}} \\
-\rho \overline{{\bf{u}}'{h'}} &=& \lambda_t \nabla \overline{T}  \\
\overline{{\bf{\Pi}}' \colon \nabla{\bf{u}}'} &=&{\bf{\Pi}}_R \colon \nabla \overline{{\bf{u}}}
\label{eq:dissfunction}
\end{eqnarray}
where ${\bf{\Pi}}_R=-\rho \overline{{\bf{u}}' \otimes {\bf{u}}'}$ is the Reynolds' stress tensor. More models are required to estimate turbulent viscosity and diffusivity, based on additional transport equations: the most well-known models are $k-\epsilon$, which solves two partial differential equations for turbulent kinetic energy and its rate of dissipation, and $k-\omega$, which uses turbulent kinetic energy and specific dissipation rate. The interested reader is referred to~\cite{Versteeg2007,Ferziger2001} for more details.

Substituting in equations (\ref{eq:RANS_ma}, \ref{eq:RANS_mo}, \ref{eq:RANS_en}):
\begin{eqnarray}
{\color{white}\frac{\partial \rho}{\partial \rho}+} \nabla \cdot \overline{\bf{u}} &=& 0  \\
\rho \frac{\partial \overline{{\bf{u}}}}{\partial t} +  \rho \nabla \cdot \left(\overline{\bf{u}} \otimes {\overline{\bf{u}}}\right) &=& -\nabla \overline{p} + \nabla \cdot \left({\overline{\bf{\Pi}}}+{\bf{\Pi}}_R\right) + \rho \overline{{\bf{a}}} \nonumber \\
&=& -\nabla \overline{p} + \left( \mu + \mu_t \right) \nabla^2 {\overline{\bf{u}}} + \rho \overline{{\bf{a}}}  \\
\rho \frac{\partial \overline{h}}{\partial t} + \rho \nabla \cdot \left({\overline{\bf{u}}} \overline{h}\right) &=& \left(\lambda + \lambda_t \right) \nabla^2 \overline{T} + \nonumber \\ 
&& + \left( \overline{{\bf{\Pi}}}+{\bf{\Pi}}_R\right) \colon \nabla{\overline{\bf{u}}} + \frac{D\overline{p}}{Dt} \nonumber \\
&=& \left( \lambda + \lambda_t \right) \nabla^2 \overline{T} +\nonumber \\ 
&&+ 2 \left( \mu+\mu_t \right) 
\left( \nabla^S {\overline{\bf{u}}} \right)^2  
+\frac{D\overline{p}}{Dt}
\end{eqnarray}

Similarly, entropy generation in a turbulent flow is affected by turbulent fluctuations, and, if these are not directly resolved by the numerical approach (i.e. in a DNS), their effect needs to be modeled. A time average of the entropy generation rate can be obtained starting from equations~(\ref{eq:sigmat},\ref{eq:sigmav}):
\begin{eqnarray}
\overline{T (\sigma_T+\sigma_V)} &=&
\overline{T} \; \overline{\sigma_T} + \overline{T' \sigma'_T} + \overline{T} \; \overline{\sigma_V} + \overline{T' \sigma'_V} \nonumber \\ 
&=& \lambda \overline{\frac{1}{T}\nabla T \cdot \nabla T} 
+ \overline{{\bf{\Pi}} \cdot {\nabla {\bf{u}}}} \nonumber \\
&=& \lambda \nabla \overline{(\text{ln} T)} \cdot \nabla \overline{T} + 
 \lambda \overline{\nabla (\text{ln} T') \cdot \nabla T'} + \nonumber \\
&& + {{\overline{\bf{\Pi}}}} \cdot {\nabla {\overline{\bf{u}}}}  
+ \overline{{\bf{\Pi'}} \cdot {\nabla {\bf{u}}'}} 
\label{eq:engen2}
\end{eqnarray}
Terms involving fluctuations of flow variables in equation (\ref{eq:engen2}) need to be modeled. Moore and Moore \cite{Moore1983,Moore1983a} suggest the following correlations for mean entropy production, thermal diffusion and viscous dissipation, respectively:
\begin{eqnarray}
\overline{T\sigma_T} &=& \frac{\lambda}{\overline{T}} \nabla \overline{T} \cdot \nabla \overline{T} + \frac{\lambda}{\overline{T}} \overline{\nabla T' \cdot \nabla T'} \\
\overline{T\sigma_V} &=&
{{\overline{\bf{\Pi}}}} \cdot \nabla {\overline{\bf{u}}}  +
 {\overline{{\bf{\Pi}}' \cdot {\nabla {\bf{u}}'}}} \\
\lambda \overline{\nabla T' \cdot \nabla T'} &=& \frac{\lambda_t}{\lambda}  \lambda \nabla \overline{T} \cdot \nabla \overline{T}  \\
\overline{\bf{\Pi'} \cdot {\nabla {\bf{u}}'}} &=& \frac{\mu_t}{\mu} {\overline{\bf{\Pi}}} \cdot {\nabla {\overline{\bf{u}}}}=  {\bf{\Pi}}_R \cdot {\nabla {\overline{\bf{u}}}}   \label{eq:gen-eq-diss}
\end{eqnarray}

Hence, in a turbulent flow, mean entropy generation rate per unit mass due to heat transfer and flow friction can be estimated with:
\begin{eqnarray}
\overline{\sigma}_T&=&\overline{\sigma}_{T,mf}+\overline{\sigma}_{T,t}\nonumber \\
&=&\frac{\lambda}{\rho \overline{T}^2}\left(\nabla \overline{T}\right)^2 + \frac{\lambda_t}{\rho \overline{T}^2}\left(\nabla \overline{T}\right)^2 \ge 0
\label{eq:sigmat_m} \\
\overline{\sigma}_V&=&\overline{\sigma}_{V,mf}+\overline{\sigma}_{V,t}\nonumber \\
&=&\frac{2(\mu)}{\rho \overline{T}}\left(\nabla^S {\overline{\bf{u}}}\right)^2 + \frac{2(\mu_t)}{\rho \overline{T}}\left(\nabla^S {\overline{\bf{u}}}\right)^2 \ge 0
\label{eq:sigmav_m}
\end{eqnarray}
In equations (\ref{eq:sigmat_m}, \ref{eq:sigmav_m}) contributions to thermal and viscous entropy production due to mean flow and turbulent fluctuations have been separated. As already noted, turbulent fluctuations in momentum and energy equations are modeled through the appearance of new stress terms (the Reynolds' stresses) and similar terms in the energy equation, modeled by replacing dynamic viscosity $\mu$ and thermal diffusivity $\lambda$ with an effective viscosity $\mu_{eff}=\mu+\mu_t$ and an effective thermal diffusivity $\lambda_{eff}=\lambda+\lambda_t$. Similarly, the terms resulting from fluctuating quantities in the entropy conservation equation can be modeled by increasing the local entropy production rate by two factors proportional to turbulent viscosity and diffusivity.

It should be noted that the above assumptions, in particular equation~(\ref{eq:gen-eq-diss}), entail a local equilibrium between production and dissipation of kinetic energy. When using RANS approaches in conjunction with two-equation turbulence models, the turbulent dissipation per unit mass is directly available (it is the $\epsilon$ of the $k-\epsilon$ model, or can be readily calculated from $k$ and $\omega$ in a $k-\omega$ method) \cite{Herwig2006}. 
While it is possible to use $\epsilon$ directly to estimate the turbulent energy dissipation ($\rho \epsilon=\overline{\bf{\Pi'} \cdot {\nabla {\bf{u}}'}}$), this would also include the dissipation of turbulent kinetic energy entering the domain, which is not linked to a loss of mean energy. The hypothesis of local equilibrium between production and dissipation of kinetic energy is an acceptable assumption, as will be shown in Section~\ref{sec:results}, and does not require a separation between the different loss sources~\cite{Paparone2003,Yamazaki2008}.

The validity of this approach has been verified by~\cite{Moore1983} for a flat plate boundary layer. The same assumption has been used by \cite{Iandoli2008,Asinari2016,Jin2015}, while different models have been used by \cite{Kock2004,Ertesvag2005,Kis2011}. {\color{black}Jin and Herwig \cite{Jin2015} showed the importance of the choice of turbulence model on the evaluation of entropy generation (or turbulent dissipation). This comes as no surprise, as the turbulence model can have a significant impact on any quantity of interested evaluated from RANS approaches. However, the focus of this paper is on the relative impact of entropy generation from mean flow and turbulence fluctuations, and on the fact that neglecting the latter can lead to significant errors and misleading conclusions \cite{Shehata2016,Shehata2017,Soltanmohamadi2016}.}

 Regardless of the specific approach, it appears clear that the contribution of turbulent dissipation to entropy production is in most cases not negligible and, unless directly found by the CFD model, it needs to be accounted for.

\section{Entropy Generation from Isolated Airfoils: the Equivalence between Entropy and Drag}
\label{sec:entropy_airfoils}

In CFD, aerodynamic forces are usually calculated by integrating pressure and viscous forces on the surface of the body of interest (\emph{near-field method}). They can also be calculated with two alternative approaches~\cite{Lock1986}:
\begin{enumerate}[label=(\alph*)]
\item from an integration on a surface surrounding the body (\emph{far-field method}), by applying the conservation of momentum to the volume enclosed between the two surfaces~\cite{Giles1999,Hunt1999,Paparone2003,Yamazaki2008};
\item from an integration in the volume enclosed between the two previously defined surfaces (\emph{mid-field method}), by application of the Gauss theorem to the integral in (a)~\cite{Paparone2003,Yamazaki2008,Chao1999,Fidkowski2012}.
\end{enumerate}
A schematic of near-field, far-field and mid-field for a typical aerodynamic problem is shown in Figure~\ref{fig:fields}.
\begin{figure}[!ht]
 \centering
 \includegraphics[width=.45\columnwidth]{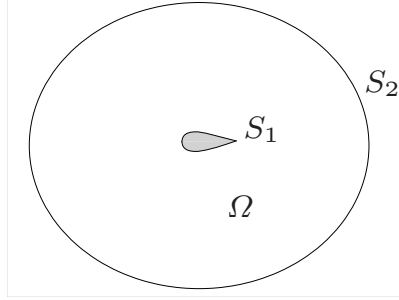}
\caption{Near-field ($S_1$), far-field ($S_2$) and mid-field ($\Omega$)}
\label{fig:fields}
\end{figure}
One of the main advantages of the mid-field method is that it allows the drag to be decomposed and visualized. The total drag can be expressed in terms of an integral of the local entropy production rate, and specific regions of the flow field can be tailored by design and optimization methods for a more effective result \cite{Yamazaki2008,Bisson2013}. 

Using the near-field method, the aerodynamic forces acting on the body represented in Figure~\ref{fig:fields} are given by:
\begin{equation}
{\bf{F}}=\int_{S_1} [(p-p_{\infty}){\bf{n}}-({\bm{\Pi}}\cdot {\bf{n}})]dS
\end{equation}
Assuming steady flow, the far-field method gives:
\begin{equation}
{\bf{F}}=-\int_{S_2} [(p-p_{\infty}){\bf{n}}-(\rho {\bf{u}}{\bf{u}}-{\bm{\Pi}}\cdot {\bf{n}})]dS
\end{equation}

If the surface $S_2$ is sufficiently far from the airfoil for the wake to have dissipated (i.e. $p\simeq p_{\infty}$ and $\bm{\Pi}=0$), the component of the force in the direction of the incoming flow (i.e. the drag $D$) can be calculated as follows:

\begin{equation}
D=-\int_{S_2} \rho u {\bf{u}} \cdot {\bf{n}}   dS 
\end{equation}

It can be shown~\cite{Paparone2003,Yamazaki2008,Fidkowski2012} that the above integral is equivalent to:

\begin{eqnarray}
D&=&\int_{S_2} u_{\infty}\left[1-\sqrt{1+\frac{2\Delta H}{u_\infty^2}+\frac{2(1-e^{\Delta s/c_p})}{(\gamma-1)M_{\infty}^2}}\right]\cdot \nonumber \\
&&\rho {\bf{u}} \cdot {\bf{n}} dS
\label{eq:drag-entropy}
\end{eqnarray}
$\Delta s$ is the specific entropy (relative to a reference value that can be taken as the value for the free-stream flow), $c_p$ is the specific heat capacity at constant pressure, $\gamma$ is the ratio of specific heats, $M_{\infty}$ is the free-stream Mach number, and $\Delta H$ is the total enthalpy relative to the free-stream value. Assuming $\Delta s/c_p<<1$ and approximating the exponential in the previous equation with a first-order expansion:

\begin{equation}
D\approx \frac{u_{\infty}}{\gamma R M_{\infty}^2}\int_{S_2} \Delta s \rho {\bf{u}} \cdot {\bf{n}} dS
\label{eq:drag}
\end{equation}

The drag $D$ is hence proportional to the net entropy flux through the far-field surface. 

The usual equation for the mid-field method can be found by considering that, in steady conditions, the net-flux of entropy through the far-field surface is equal to the entropy generated per unit time inside the mid-field volume $\Omega$ ($\dot{S}_G$):

\begin{equation}
\int_{S_2} \Delta s \rho {\bf{u}} \cdot {\bf{n}} dS = \int_{\Omega} \rho (\sigma_T+\sigma_V) d\Omega = \dot{S}_G
\label{eq:sg}
\end{equation}
where $\sigma_T$ and $\sigma_V$ are entropy generation rates per unit mass due to heat transfer and viscous dissipation, respectively (see equations (\ref{eq:sigmat}, \ref{eq:sigmav}), or (\ref{eq:sigmat_m}, \ref{eq:sigmav_m}), for a turbulent flow). 
Even though dynamic effects are negligible for the cases considered in this work (this will be discussed in detail in Section~\ref{sec:results_unsteady}), it should be noted that the mid-field method (also known as drag-decomposition or breakdown method) has recently been extended to unsteady flows \cite{Gariepy2013,Toubin2016}.

In addition, considering that $M_{\infty}=u_{\infty}/\sqrt{\gamma R T_{\infty}}$ and that the drag can be expressed via the drag coefficient $c_d$, equation (\ref{eq:drag}) can be expressed as:
\begin{equation}
c_d=\frac{D}{\frac{1}{2}\rho u_{\infty}^2 A}
\label{eq:cd}
\end{equation}
\begin{equation}
D=(\frac{1}{2}\rho u_{\infty}^2 A) c_d =\frac{T_{\infty}}{u_{\infty}} \dot{S}_G
\label{eq:d-sg}
\end{equation}

A non-dimensional entropy generation coefficient $K_{\dot{S}_G}$ can be defined as follows:

\begin{equation}
K_{\dot{S}_G}=\frac{T_{\infty}\dot{S}_G}{\frac{1}{2}\rho u_{\infty}^3 A}
\label{eq:ksg}
\end{equation}

From equation~(\ref{eq:d-sg}) and the above definition
\begin{equation}
c_d =K_{\dot{S}_G}
\label{eq:cd-ksg}
\end{equation}
Under the above assumptions the non-dimensional global entropy production rate $K_{\dot{S}_G}$ is simply equal to the drag coefficient $c_d$. In the following sections, the validity of equation~(\ref{eq:cd-ksg}) will be assessed for an isolated NACA0012 profile, before applying this methodology to quantify the losses in a Wells turbine.

\section{Numerical Simulations}
\label{sec:methods}

Numerical simulations have been performed to analyze entropy generation in Wells turbines, and in particular the relative importance of entropy produced by mean flow and turbulent fluctuations. These results have been compared to those presented by~\cite{Shehata2016,Soltanmohamadi2016}, who calculated entropy generation in Wells turbines, neglecting the contribution of turbulent fluctuations and reaching plainly unphysical conclusions, such as the reduction of entropy production (hence losses) with incidence angle. 

In addition, since the work of~\cite{Shehata2016} simulates a Wells turbine as an isolated airfoil (perhaps with a questionable simplification), the numerical analysis will be extended to a similar case. The comparison with the results of~\cite{Shehata2016}  provides an opportunity to understand the impact of turbulence on the entropy production in a problem where it can be readily linked to aerodynamic drag, a concept aerodynamicists and researchers are probably more familiar with.

{\color{black}The $k-\omega$ turbulence model has been used in this work, as it performed better than other models in the prediction of forces for the Wells turbine under evaluation \cite{Ghisu_JFE}. It is important to highlight that any other model could have been chosen, and would have produced similar results, as the focus of this work is the evaluation of the relative importance of the entropy generation rates produced by mean flow and turbulent fluctuations, rather than a precise evaluation of the overall entropy generation rate.}

It is also worth noting that all results have been calculated using steady-state simulations. This has been possible due to the negligible contribution of dynamic effects for both isolated airfoils~\cite{Carr1977,McCroskey1981} and Wells turbines~\cite{Ghisu_JTS,Ghisu_JFE,Ghisu_JoPaE}, at the reduced frequencies Wells turbines typically operate at ($k=(\pi f c)/u_{\infty}< 10^{-3}$). {\color{black}This point will be discussed in detail in Section~\ref{sec:results_unsteady}.}

For all simulations, forces (and hence force coefficients) and entropy generation rates have always been calculated \emph{directly}. This means that forces have been derived by integrating pressure and viscous forces on blade surfaces, while entropy generation rates have been found using a volume integral of local values of $\overline{\sigma}_v$ (see equations (\ref{eq:sigmav}) and (\ref{eq:sigmav_m})). Due to the assumption of adiabatic incompressible flow, contributions due to heat transfer (equation (\ref{eq:sigmat}), or (\ref{eq:sigmat_m})) have been neglected. The analogy between entropy and drag (see Section~\ref{sec:entropy_airfoils}) has then been used to verify the approach.
In addition, entropy generation rate $\dot{S}_G$, defined in equation (\ref{eq:sg}), has been divided into the two contributions due to mean flow ($\dot{S}_{G,mf}$) and turbulent fluctuations ($\dot{S}_{G,t}$), as follows:
\begin{eqnarray}
\dot{S}_{G,mf}&=&\int_{\Omega} \rho (\overline{\sigma}_{V,mf}) d\Omega =\int_{\Omega} \frac{2\mu}{T} (\nabla^S {\overline{\bf{u}}})^2 d\Omega \\
\dot{S}_{G,t}&=&\int_{\Omega} \rho (\overline{\sigma}_{V,t}) d\Omega =\int_{\Omega} \frac{2\mu_t}{T} (\nabla^S {\overline{\bf{u}}})^2 d\Omega \\
\dot{S}_{G}&=&\dot{S}_{G,mf}+\dot{S}_{G,t}
\end{eqnarray}

Similarly, the non-dimensional entropy generation rate $K_{\dot{S}_G}$, defined in equation (\ref{eq:ksg}), has been divided into the contributions due to mean flow and turbulent fluctuations ($K_{\dot{S}_{G,mf}}$ and $K_{\dot{S}_{G,t}}$, respectively):

\begin{eqnarray}
K_{\dot{S}_{G,mf}}&=& \frac{T_{\infty}\dot{S}_{G,mf}}{\frac{1}{2}\rho u_{\infty}^3 A} \label{eq:ksg1}\\
K_{\dot{S}_{G,t}}&=& \frac{T_{\infty}\dot{S}_{G,t}}{\frac{1}{2}\rho u_{\infty}^3 A} \label{eq:ksg2}\\\\
K_{\dot{S}_{G}}&=& K_{\dot{S}_{G,mf}}+K_{\dot{S}_{G,t}} \label{eq:ksg3}
\end{eqnarray}

\subsection{Part I: Isolated NACA0012 Airfoil}

Simulations on an isolated NACA0012 airfoil have been performed with ANSYS Fluent, using a $k-\omega$ SST turbulence model with low-Reynolds-correction.
The SIMPLEC algorithm has been used for pressure-velocity coupling, a second-order upwind scheme for discretizing convective terms and a second-order centered scheme for pressure and viscous terms. 
 A schematic of the computational domain is shown in Figure~\ref{fig:domain_airfoil}, together with the conditions set at the boundaries, which have been placed at a distance of 15 chords from the airfoil. A multi-block structured mesh has been used, with particular attention to the boundary layer region (the maximum non-dimensionalized wall distance $y^+$ is of the order of 1). The detail of the mesh near the airfoil is reported in Figure~\ref{fig:domain_airfoil}. 

\begin{figure}[!ht]
\centering
\includegraphics[width=.8\columnwidth]{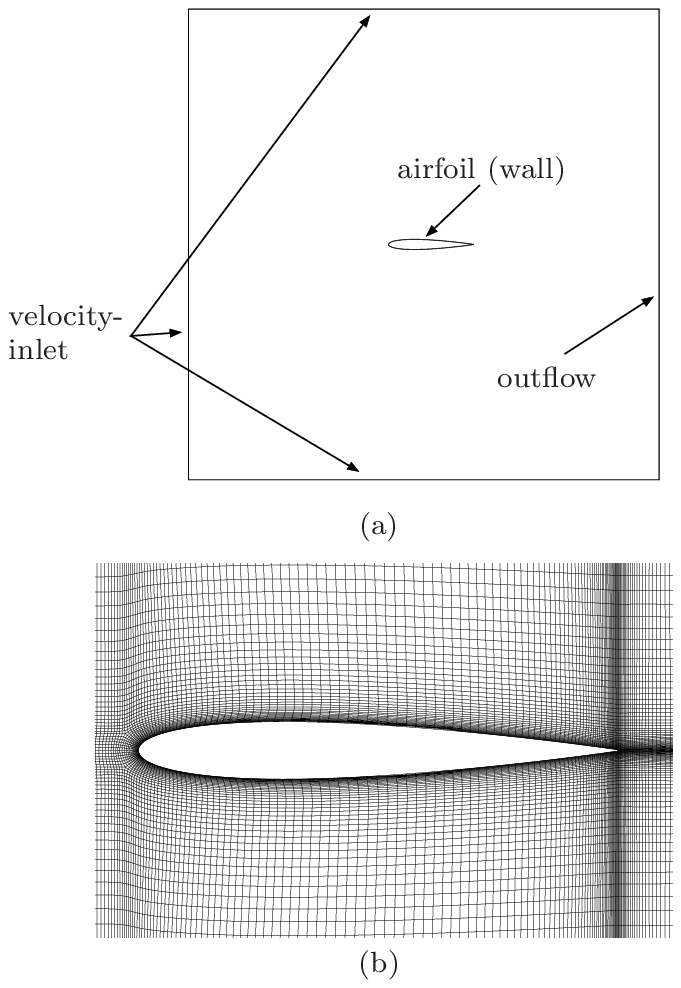}
\label{fig:domain_airfoil}
\caption{Computational domain (a, not to scale) and numerical mesh (b, every third grid line)}
\end{figure}

Steady simulations have been performed, for different angles of attack and Reynolds numbers, while keeping the Mach number fixed ($M_{\infty}=0.1$), to preserve the validity of the incompressibility assumption. 

\subsection{Part II: Wells Turbine}

The Wells turbine of Setoguchi \emph{et al.}~\cite{Setoguchi1998a} has been studied using the same geometrical simplification of~\cite{Setoguchi2003b}, recently adopted also by \cite{Soltanmohamadi2016}. This means that only the duct that houses the turbine is modeled, as reported in Figure~\ref{fig:domain_wells}, together with the boundary conditions used. {\color{black}A summary of the main geometric parameters and operating conditions is reported in Table~\ref{tab:seto}.

\begin{table}[!h]
\centering
\caption{Geometrical and operating data for Setoguchi's experiments \cite{Setoguchi1998}}\label{tab:seto}
\begin{tabular}{lc}
\hline
			chamber diameter [m] &  1.4 \\
			rotor tip diameter [mm] &  300 \\
			rotor hub diameter [mm] &  210\\
			tip clearance [mm] &  1\\
			chord length $c$ [mm] &  90\\
			sweep ratio [-] &  0.417 \\
			number of blades [-]&   6  \\
			blade profile  & NACA0020 \\
			solidity at tip radius $\sigma$ [-]&  0.57 \\
			rotational speed [rpm]&  2500\\
			operating frequency $f$ [s$^{-1}$]&  $1/6$\\
			Reynolds number $Re$ [-]&  $2 \times 10^5$\\
			Mach number $M$ [-]&  0.1\\
			turbine non-dimensional frequency $k$ [-]&  0.0014\\
\hline
\end{tabular}
\end{table}
}

A multi-block structured grid (Figure \ref{fig:mesh_wells}) was used to discretize the governing equations. A C-grid around the blade was able to capture the complex boundary layer flow, with an H-mesh structure in the rest of domain. 
A grid sensitivity study was conducted to verify the choice of the numerical mesh. For the basic mesh, 260 points were used around the blade profile, and 70 between successive blades (in the wake region). In the spanwise direction, 35 points cover the blade span, while 10 points were employed in the tip gap region, for a total of about $10^6$ cells.  The maximum $y^+$ is of the order of 1 to ensure a good resolution of the boundary layer. 

Steady-state simulations were performed for different values of the flow coefficient $\phi$, defined in equation~(\ref{eq:phi}). 
 The $k-\omega$ SST model was selected for turbulence closure. 
The SIMPLEC algorithm was used for pressure-velocity coupling, a second-order upwind scheme for discretizing convective terms and a second-order centered scheme for pressure and viscous terms. Multiple reference frames allow the interaction between stationary and rotating regions to be simulated. More details about validation and verification, including turbulence closure and dynamic effects, can be found in \cite{Ghisu_JFE}.

\begin{figure}[!h]
\begin{center}
\includegraphics[width=.4\textwidth]{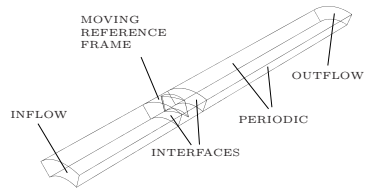}
\caption{Computational domain for Wells turbine simulations}
\label{fig:domain_wells}
\end{center}
\end{figure}

\begin{figure}[!h]
        \centering
        \includegraphics[clip,width=0.36\textwidth]{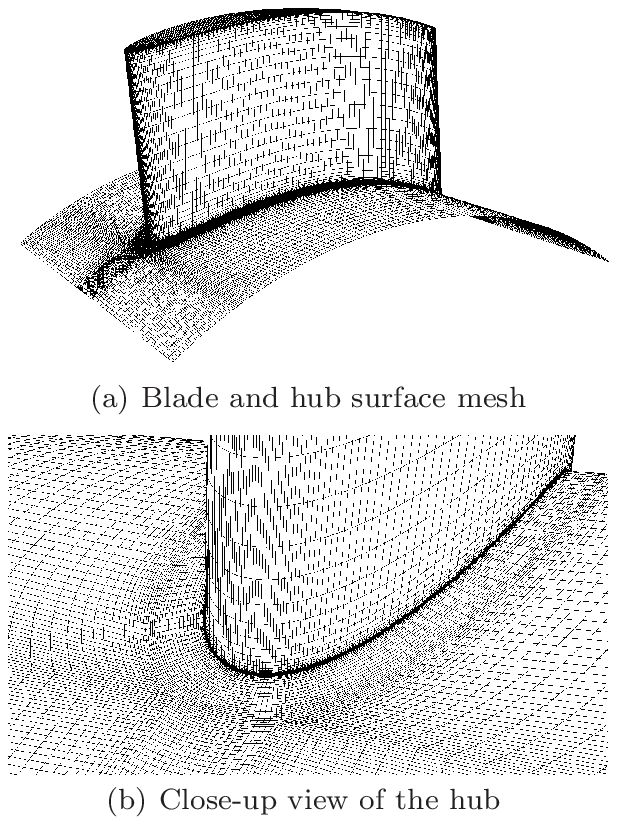}
\caption{Details of the computational mesh for Wells turbine simulations}
\label{fig:mesh_wells}
\end{figure}

\section{Results}
\label{sec:results}

\subsection{Part I: Isolated Airfoil}
\label{sec:results_airfoil}

The drag coefficient $c_d$ and the non-dimensional entropy generation rate $K_{\dot{S}_G}$ for the NACA0012 profile are presented in Figure~\ref{fig:cdksg-alpha}, as a function of the incidence angle, for different Reynolds numbers. 
The $c_d-\alpha$ plots follow well-known trends (see \cite{Abbott_TWS} as an example): for a given Reynolds number, the drag coefficient $c_d$ grows steadily with the angle of attack, up to a critical angle, when a dramatic rise is caused by the rapid increase in boundary layer thickness leading to a large separation. Increasing the Reynolds number has two effects: it reduces boundary layer thickness (and hence $c_d$) and, thanks to a more energetic boundary layer, it postpones the occurrence of separation, increasing the value of the critical angle of attack. 

The $K_{\dot{S}_G}-\alpha$ plots follow the same trends. Aside from proving the theoretical results presented in Sections~\ref{sec:entropy}~and~\ref{sec:entropy_airfoils}, this also demonstrates the correctness of the numerical approach, as well as the validity of hypothesis of local equilibrium between production and dissipation of turbulent kinetic energy. Small differences can be attributed to the simplifications adopted (e.g. the size of the domain) or to numerical errors (most numerical schemes do not conserve entropy, hence the entropy downstream of the airfoil can be polluted by spurious entropy generation~\cite{Fidkowski2012}). However, discrepancies are never larger than a few percent.

\begin{figure}[!ht]
 \centering
 \includegraphics[trim={0 0 0 .0},clip,width=.75\columnwidth]{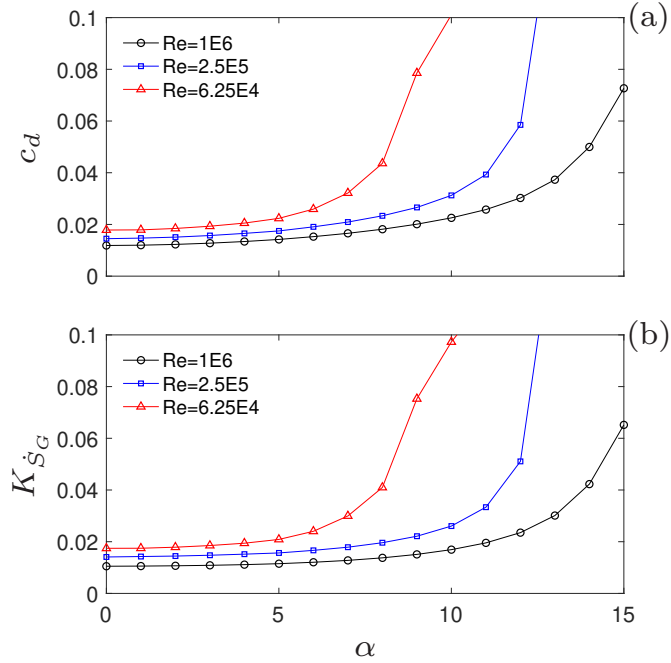}
\caption{Dependence of drag coefficient $c_d$ (a) and non-dimensional global entropy generation rate $K_{\dot{S}_G}$ (b) on the incidence angle, for different Reynolds numbers}
\label{fig:cdksg-alpha}
\end{figure}

In Figure \ref{fig:ksg_contributions}, the non-dimensional entropy generation rate $K_{\dot{S}_G}$ has been subdivided into the two contributions due to mean flow and turbulent fluctuations ($K_{\dot{S}_{G,mf}}$ and $K_{\dot{S}_{G,t}}$). Results are reported as a function of the incidence angle, for three values of Reynolds number. The relative contribution of turbulent fluctuations increases both for larger Reynolds numbers and larger incidence angles, due to the larger turbulent dissipation that is present in these conditions. Even at the lowest value of Reynolds number studied here (6.25e4), the contribution of turbulent fluctuations cannot be neglected: at low angles of attack, it is about 1/3 of the total, at 6$^{\circ}$ it equals the contribution of mean flow, and grows dramatically for larger incidence angles (due to the appearance of extensive separation regions). Neglecting the contribution of turbulent fluctuations leads not only to incorrect values of entropy generation, but more importantly to misleading conclusions, such as that an airfoil at very large incidence angles produces less entropy than the same airfoil at low angles of attack (as incorrectly reported in \cite{Shehata2016}).  
It is clear, from the results presented in Figure \ref{fig:ksg_contributions}, that an optimization that accounts only for the entropy generated from the mean flow can lead to wrong results.
\begin{figure}[!ht]
 \centering
 \includegraphics[trim={0 0 0 .0},clip,width=.75\columnwidth]{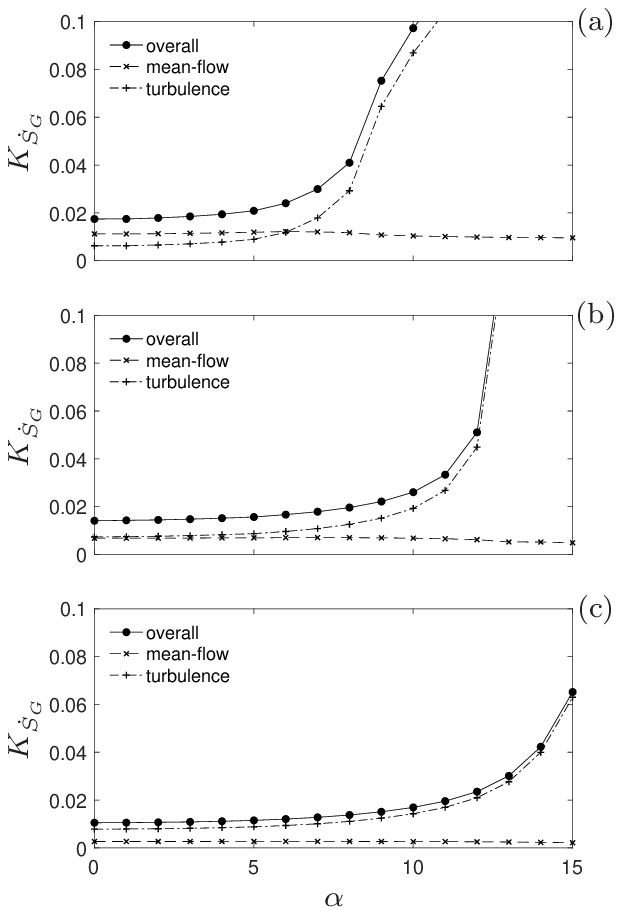}
\caption{Contributions of mean flow and turbulent fluctuations to entropy generation, as a function of the incidence angle, for different Reynolds numbers: (a) $Re=6.25E4$, (b) $Re=2.5E5$, (c) $Re=1E6$}
\label{fig:ksg_contributions}
\end{figure}

Figure~\ref{fig:cdksg-re} presents the dependence of $c_d$ and $K_{\dot{S}_G}$ on the Reynolds number, for different incidence angles. For attached flows ($\alpha=0^{\circ}$ and $\alpha=2^{\circ}$), $c_d$ and $K_{\dot{S}_G}$ exhibit a slow reduction, that becomes more pronounced at higher angles of attack (where the lowest Reynolds number causes the airfoil to stall). 

The same results are also presented in dimensional form in Figures~\ref{fig:sg-re} and \ref{fig:sg-re2}. Even though this makes results less readily comparable and does not follow best practice, some authors~\cite{Shehata2016} have presented entropy generation rates for airfoils in this form, perhaps thereby concealing important errors. Figure~\ref{fig:sg-re} presents the entropy generation rate $\dot{S}_G$ as a function of the Reynolds number, for a given airfoil. $\dot{S}_G$ increases rapidly, as it is proportional to the third power of $Re$, or of $u_{\infty}$, for a given geometry. Separating the contributions of mean flow and turbulent fluctuations, as in Figure~\ref{fig:sg-re2}, highlights again how considering only the contribution of mean flow would lead to a severe underestimation of the rate of increase of entropy generation with Reynolds number, or flow velocity.

\begin{figure}[!ht]
 \centering
 \includegraphics[trim={0 0 0 .0},clip,width=.75\columnwidth]{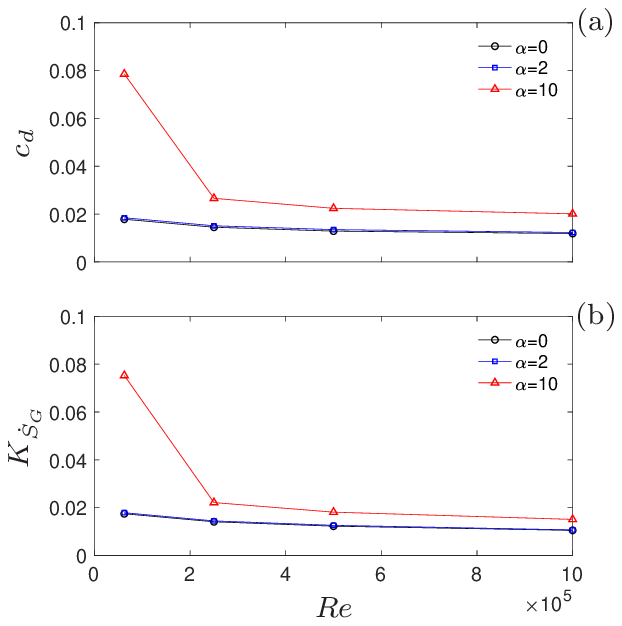}
\caption{Dependence of drag coefficient $c_d$ (a) and non-dimensional global entropy generation rate $K_{\dot{S}_G}$ (b) on Reynolds number, for different incidence angles}
\label{fig:cdksg-re}
\end{figure}

\begin{figure}[!ht]
 \centering
 \includegraphics[trim={0 0 0 .0},clip,width=.75\columnwidth]{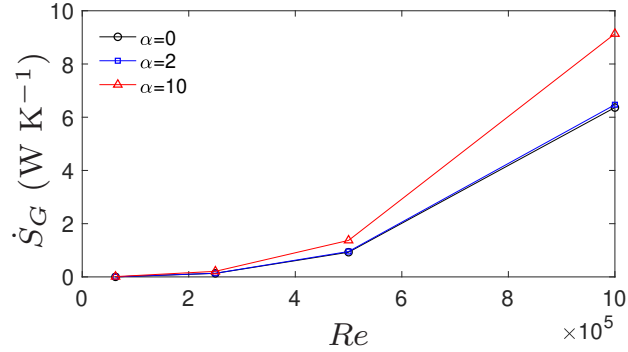}
\caption{Dependence of global entropy generation rate $\dot{S}_G$ on Reynolds number, for different incidence angles}
\label{fig:sg-re}
\end{figure}

\begin{figure}[!ht]
 \centering
 \includegraphics[trim={0 0 0 .0},clip,width=.75\columnwidth]{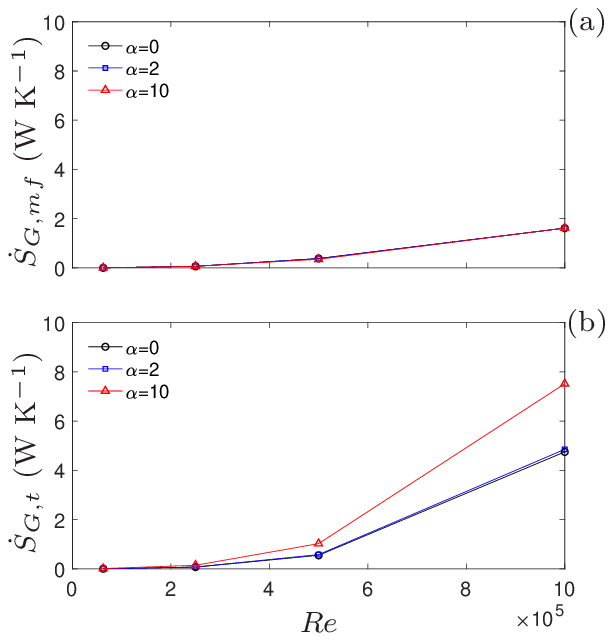}
\caption{Dependence of global entropy generation rates due to mean flow (a) and to turbulent fluctuations (b) on Reynolds number, for different incidence angles}
\label{fig:sg-re2}
\end{figure}

The identification of local entropy generation rates can be important to isolate the main loss sources, and to improve performance.
Figures \ref{fig:sigmacontours_Re6e4}, \ref{fig:sigmacontours_Re2e5} and \ref{fig:sigmacontours_Re1e6} present contours of  entropy generation rates per unit mass, separated into the contributions of mean flow (left) and turbulent fluctuations (right) for different incidence angles and Reynolds numbers. 
These terms have been made non-dimensional, to be more easily comparable and reproducible. 
The logarithmic scale allows the visualization of features of different orders of magnitude. 
The contribution of turbulent fluctuations is important, even for the lowest Reynolds number and for attached flows. Increasing Reynolds number and angle of attack leads to large turbulent dissipation regions and areas of separated flows are linked to a dramatic rise in entropy production. This cannot be predicted by considering only the contribution of the mean flow, as turbulent fluctuations are not solved in a RANS approach.

\begin{figure}[!ht]
 \centering
 \includegraphics[trim={0 0 0 .0},clip,width=.94\columnwidth]{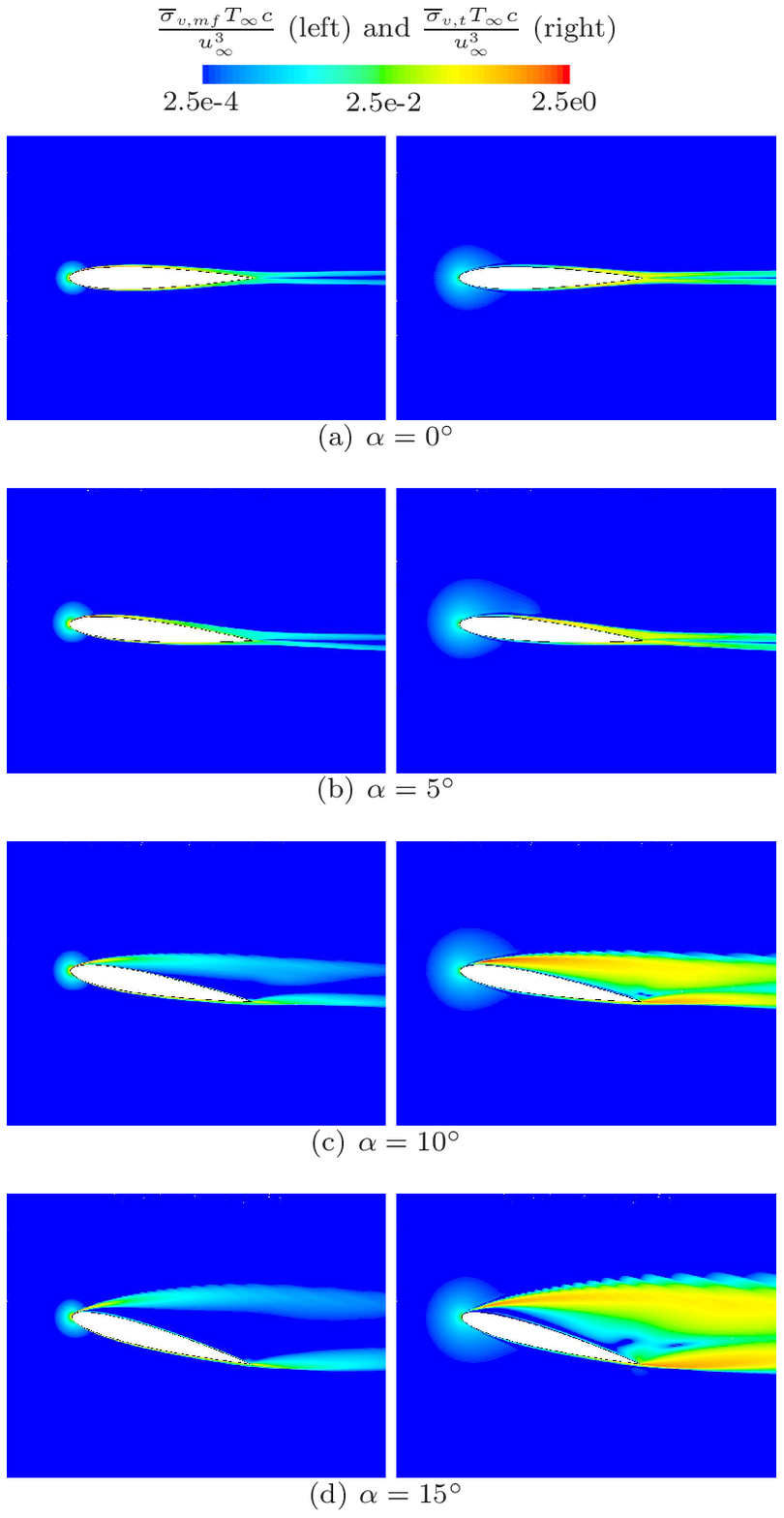}
\caption{Non-dimensionalized local entropy production rates for a NACA0012 airfoil at various incidence angles, Re=6.25e4. On the left the contribution due to mean flow, on the right the contribution of turbulent fluctuations}
\label{fig:sigmacontours_Re6e4}
\end{figure}

\begin{figure}[!ht]
 \centering
 \includegraphics[trim={0 0 0 .0},clip,width=.94\columnwidth]{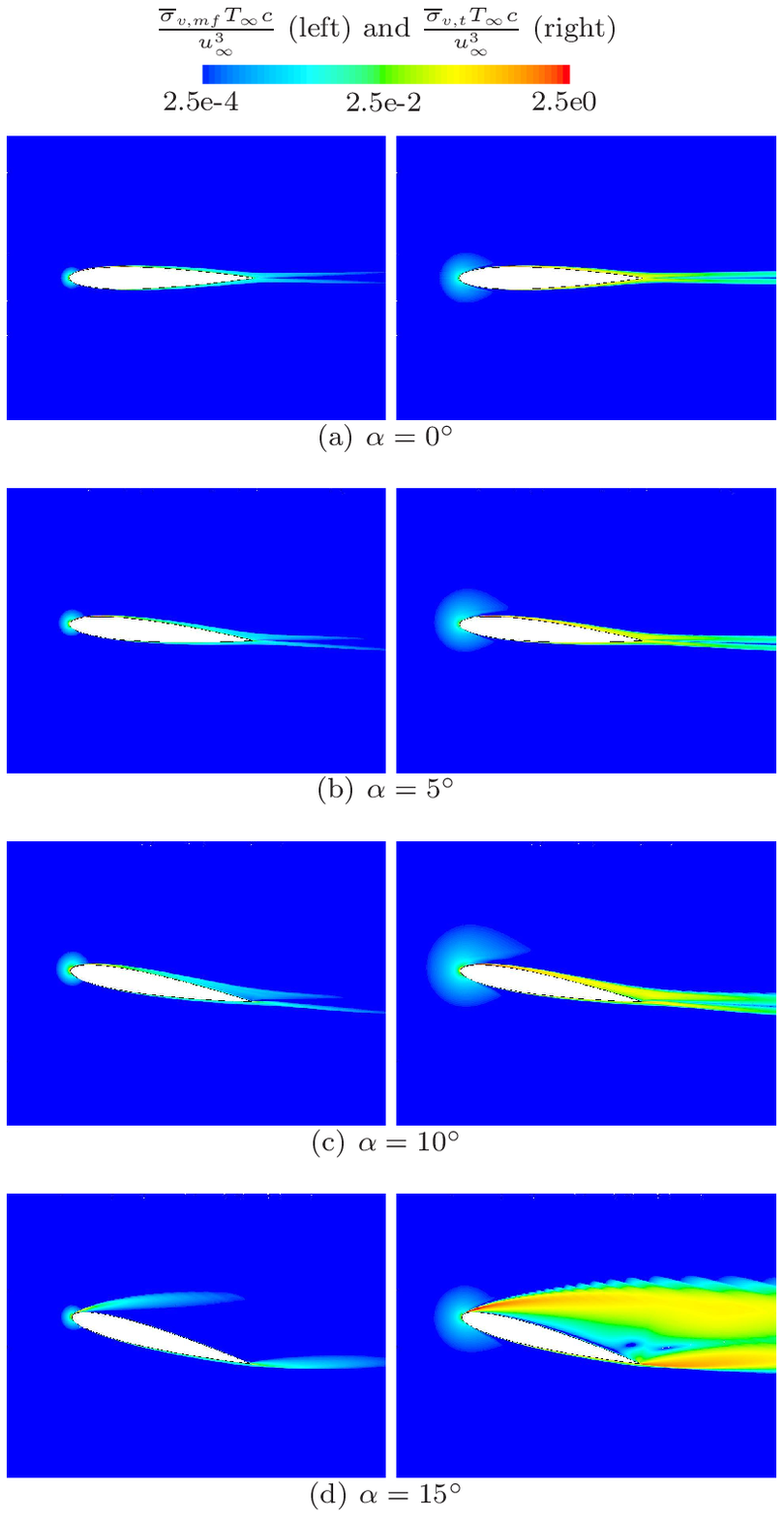}
\caption{Non-dimensionalized local entropy production rates for a NACA0012 airfoil at various incidence angles, Re=2.5e5. On the left the contribution due to mean flow, on the right the contribution of turbulent fluctuations}
\label{fig:sigmacontours_Re2e5}
\end{figure}

\begin{figure}[!ht]
 \centering
 \includegraphics[trim={0 0 0 .0},clip,width=.94\columnwidth]{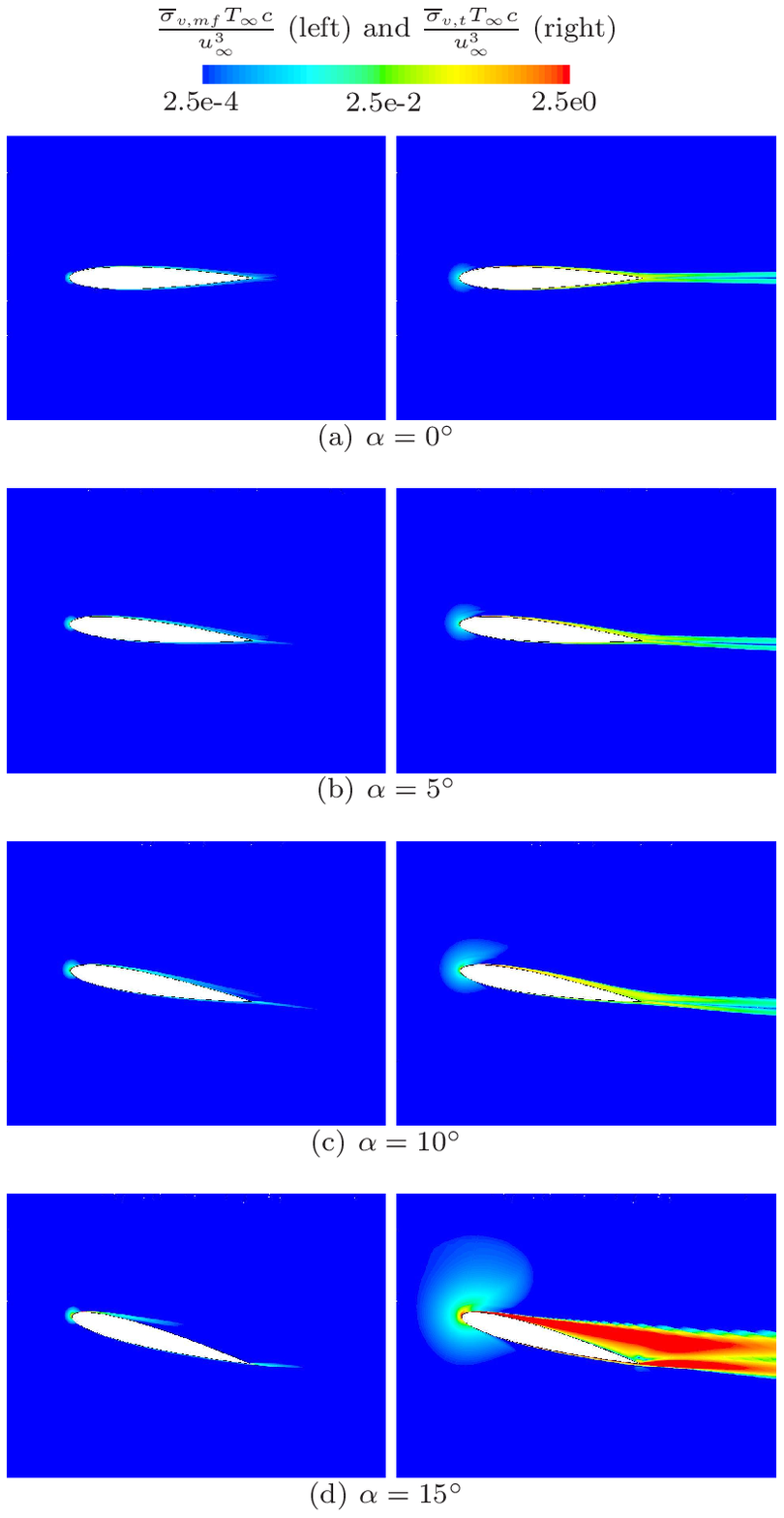}
\caption{Non-dimensionalized local entropy production rates for a NACA0012 airfoil at various incidence angles, Re=1e6. On the left the contribution due to mean flow, on the right the contribution of turbulent fluctuations}
\label{fig:sigmacontours_Re1e6}
\end{figure}

\subsection{Part II: Wells Turbine}
\label{sec:results_wells}

Figure~\ref{fig:torque-deltap} presents the results for the simulation of the Wells turbine of Setoguchi~\emph{et al.}~\cite{Setoguchi1998a}, in terms of non-dimensional torque $T^*$ and static pressure drop $P^*$, as a function of the flow coefficient $\phi$, defined as follows:
\begin{eqnarray}
T^*=\frac{T}{\rho\omega^2 r_{tip}^5}; \;\;\;
P^*=\frac{\Delta P}{\rho\omega^2 r_{tip}^2}; \;\;\;
\phi=\frac{V_a}{\omega r_{tip}}
\label{eq:phi}
\end{eqnarray}
where $T$ is the torque, $\Delta P$ the pressure drop across the rotor, $V_a$ is the mean axial flow velocity upstream of the rotor, $\omega$ its rotational speed and $r_{tip}$ its tip radius.
The experimental data present a hysteresis due to the capacitive behavior of the OWC system, discussed in detail in~\cite{Ghisu_JFE}. The numerical results (at different flow coefficients within the operating range) reproduce the experimental behavior satisfactorily. More details about mesh independence, turbulence closure, and dynamic effects can be found in~\cite{Ghisu_JFE} and are not reported here.

\begin{figure}[!ht]
 \centering
 \includegraphics[trim={0 0 0 .0},clip,width=.75\columnwidth]{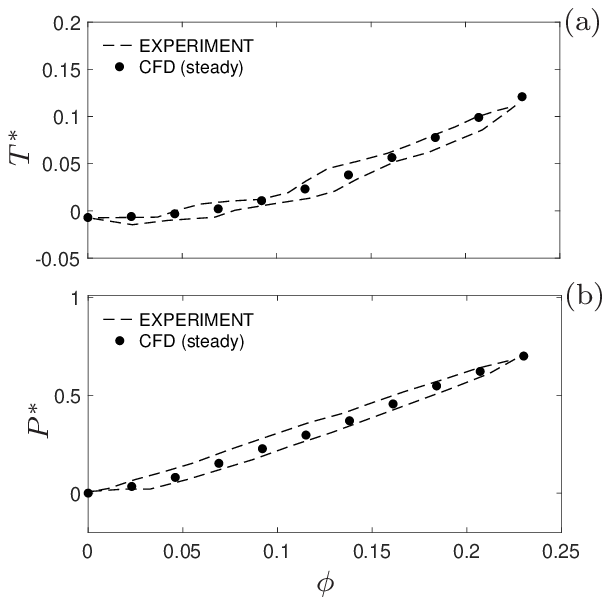}
\caption{Wells turbine performance in terms of non-dimensional torque $T^*$ (a) and static pressure drop $P^*$ (b) as a function of flow coefficient $\phi$}
\label{fig:torque-deltap}
\end{figure}

Figure~\ref{fig:wells_entropy} reports the non-dimensional global entropy generation rate $K_{\dot{S}_G}$ as a function of flow coefficient $\phi$, together with the contributions given by mean flow ($K_{\dot{S}_{G,mf}}$) and turbulent fluctuations ($K_{\dot{S}_{G,t}}$). The blade tip speed has been used as the reference velocity to non-dimensionalize global entropy generation rate values (equations~(\ref{eq:ksg1}, \ref{eq:ksg2}, \ref{eq:ksg3})).

It is clear that the entropy production increases rapidly for larger flow coefficients, for the same reasons already explained in the case of the isolated airfoil,  as the flow coefficient for a Wells turbine is a measure of the incidence angle ($\alpha=\text{tan}^{-1}(\phi)$). Similarly, in this case also, the contribution of the mean flow to the global entropy production is limited, even for low values of $\phi$ (about one third of the total). At larger flow coefficients, the contribution of turbulent fluctuations dominates, becoming about 85\% at $\phi=0.23$. 

\begin{figure}[!ht]
 \centering
 \includegraphics[trim={0 0 0 .0},clip,width=.75\columnwidth]{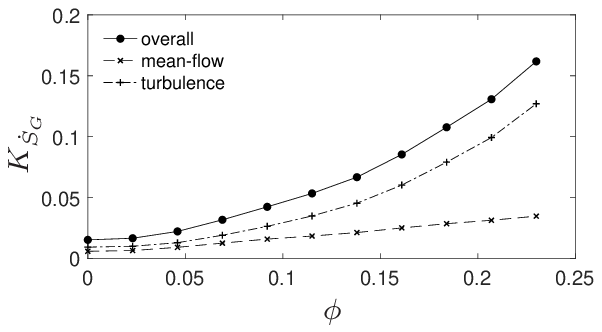}
\caption{Non-dimensional global entropy generation rate as a function of flow coefficient $\phi$}
\label{fig:wells_entropy}
\end{figure}

Figures~\ref{fig:radialcont_phi12} and \ref{fig:radialcont_phi23} present contours of non-dimensional local entropy generation rates at different locations along the blade span. This analysis can provide important information about the main sources of entropy generation. Contributions due to mean flow and to turbulent fluctuations are presented, and the latter appear significantly larger at every spanwise position. The appearance of a large region of separated flow near the tip of the blade, characterized by a large entropy generation rate, is evident, together with the effect of the wake generated by preceding blades. Neglecting the contribution of turbulent fluctuations to entropy generation could lead to misleading conclusions, such as those reported in~\cite{Soltanmohamadi2016}, where entropy generation (and therefore losses) do not increase for larger flow coefficients. 

\begin{figure}[!h]
\begin{center}
\includegraphics[trim={0 0 0 .0},clip,width=.94\columnwidth]{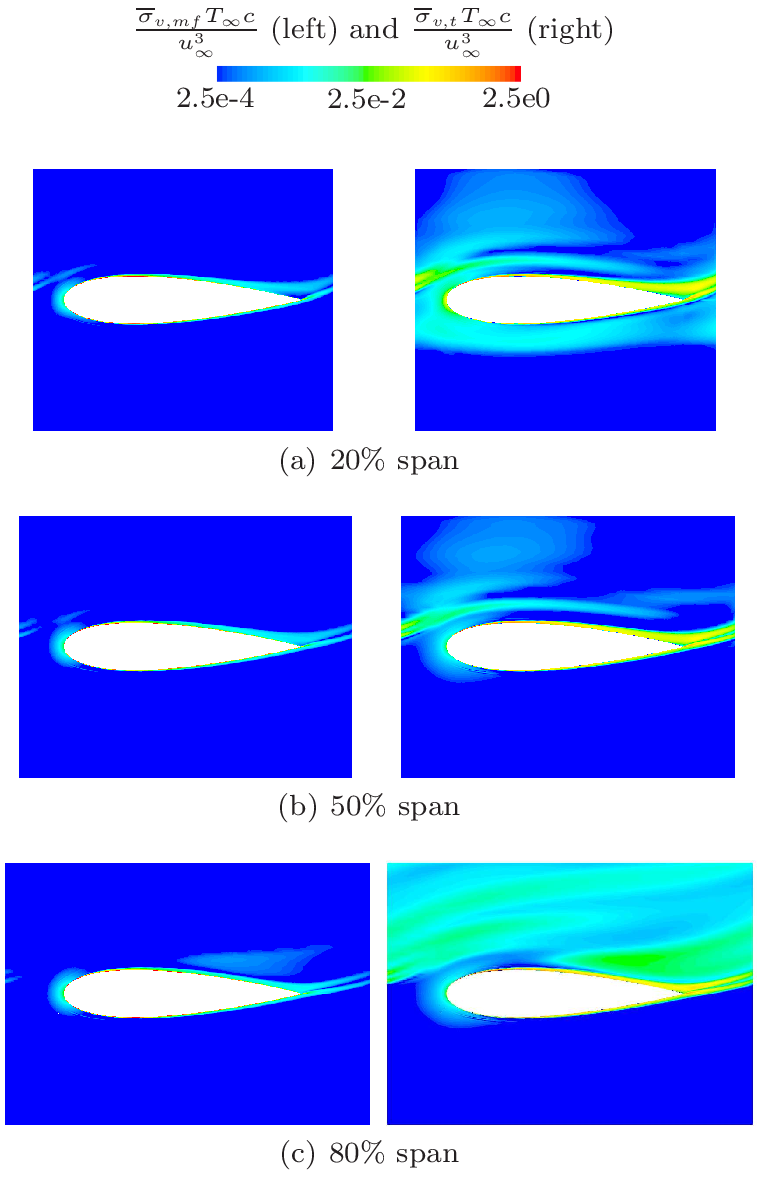}
\caption{Non-dimensional local entropy generation rates due to mean flow (left) and turbulence (right), at 3 radial positions ($\phi=0.12$)}
\label{fig:radialcont_phi12}
\end{center}
\end{figure}

\begin{figure}[!h]
\begin{center}
\includegraphics[trim={0 0 0 .0},clip,width=.94\columnwidth]{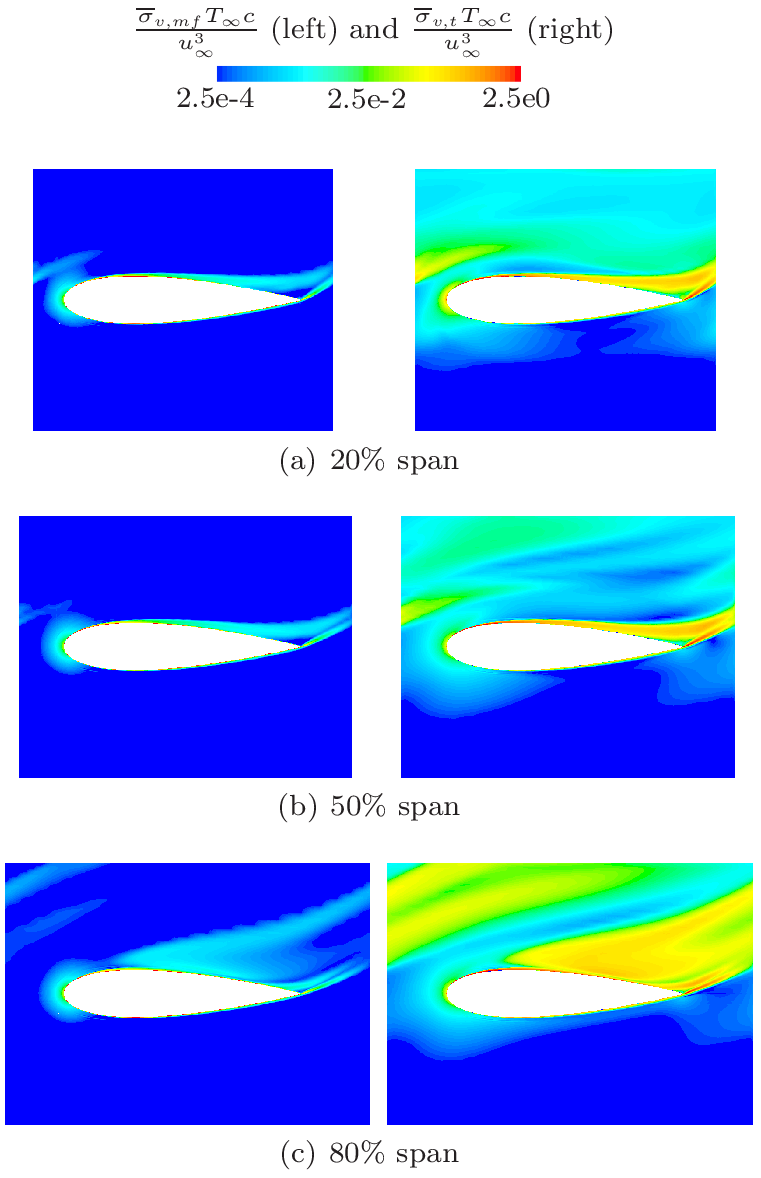}
\caption{Non-dimensional local entropy generation rates due to mean flow (left) and turbulence (right), at 3 radial positions ($\phi=0.23$)}
\label{fig:radialcont_phi23}
\end{center}
\end{figure}

Figures~\ref{fig:tangentialcont_phi12} and~\ref{fig:tangentialcont_phi23} present the same results on planes located at different tangential positions (perpendicular to the machine axis). This analysis highlights the main sources of entropy generation, such as boundary layer and secondary flows (tip vortex and horseshoe vortices). The strength of the tip vortex grows for larger flow coefficients and moving towards the trailing edge of the blade, and the presence of the tip vortex generated by the preceding blade is also evident. The horseshoe vortex, generated near the leading edge, loses its strength towards the trailing edge. The boundary layer thickness grows due to the adverse pressure gradient on the blade, and for the larger flow coefficient ($\phi=0.23$) it appears on the verge of separation. The wake generated by the preceding blade is also evident, at least for the lower value of flow coefficient ($\phi=0.12$), while the situation becomes more confused at the highest $\phi$, due to the larger wake that covers most of the scene and the mixing with other large secondary flow structures. All these flow features appear significantly less evident when looking at the contribution from the mean flow, which, as mentioned, is almost negligible when compared with the contribution of turbulent fluctuations.

\begin{figure}[!ht]
\begin{center}
\includegraphics[trim={0 0 0 .0},clip,width=.94\columnwidth]{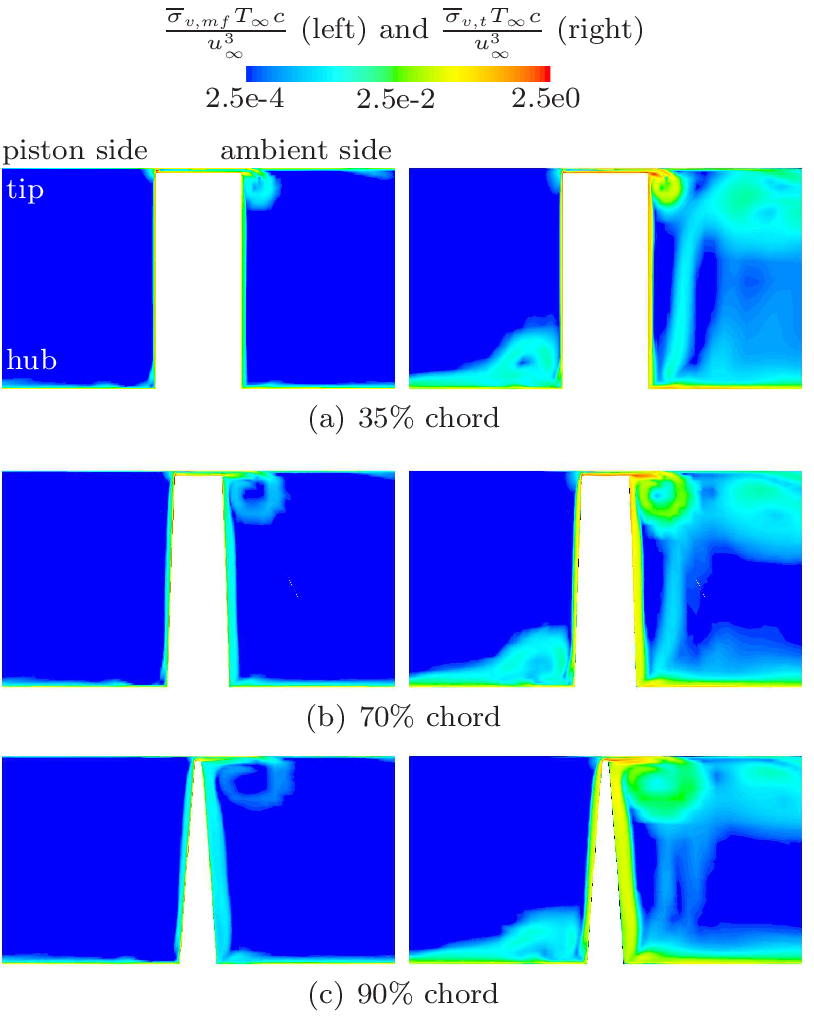}
\caption{Non-dimensional local entropy generation rates due to mean flow (left) and turbulence (right), at 3 tangential positions ($\phi=0.12$)}
\label{fig:tangentialcont_phi12}
\end{center}
\end{figure}

\begin{figure}[!ht]
\begin{center}
\includegraphics[trim={0 0 0 .0},clip,width=.94\columnwidth]{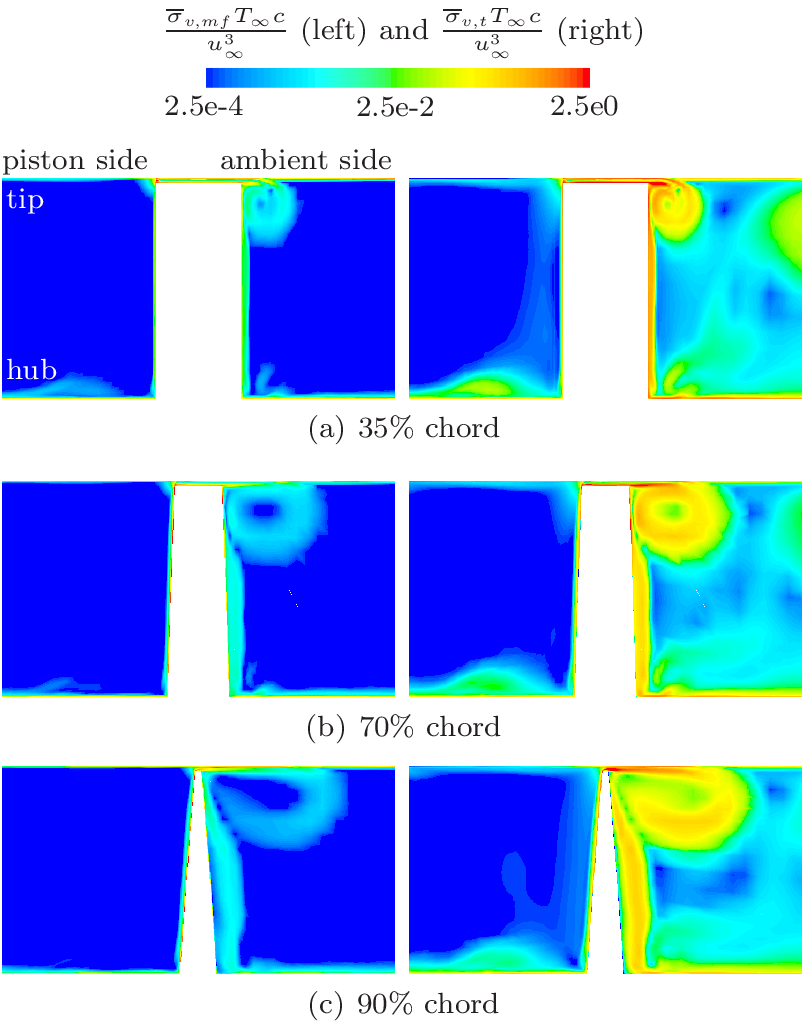}
\caption{Non-dimensional local entropy generation rates due to mean flow (left) and turbulence (right), at 3 tangential positions ($\phi=0.23$)}
\label{fig:tangentialcont_phi23}
\end{center}
\end{figure}

{\color{black}
\subsection{Part III: Unsteady Effects}
\label{sec:results_unsteady}

The results presented in Sections \ref{sec:results_airfoil} and \ref{sec:results_wells} were obtained with steady CFD simulations. As mentioned in Section \ref{sec:methods}, a Wells turbine operates at non-dimensional frequencies lower than 10$^{-3}$, not sufficient to generate appreciable dynamic effects \cite{Ghisu_JFE}.

Several authors have discussed the presence of hysteresis in Wells turbines \cite{Kinoue2003,Setoguchi2003b,Shehata2016,Shehata2017,Hu2018}. However, the authors of this article have shown in previous works how the alleged aerodynamic hysteresis was caused by an incorrect temporal discretization \cite{Ghisu_JFE,Ghisu_JoPaE}: the real cause of the hysteresis shown in the experimental analyses \cite{Setoguchi1998} is the capacitive behavior of the OWC system \cite{Ghisu_JFE,GhisuAIAAJ}.

To reinforce these statements, unsteady simulations have been performed for the geometry presented in Figure~\ref{fig:domain_wells}. The velocity at the inflow boundary was set as a periodic function of time, to produce a sinusoidal variation of the flow coefficient, with a period of 6 s, corresponding to a non-dimensional frequency $k=1.4 \times 10^{-3}$. Three periods were simulated to ensure periodically stable reults.

The unsteady RANS equations were solved using a first-order implicit method for the temporal discretization, with a time step of 10$^{-4}$ and 20 sub-iterations, chosen after a careful sensitivity analysis to avoid the presence of spurious phase delays, which would result in a false hysteresis. More information on the complete verification and validation process is reported in ~\cite{Ghisu_JFE}. 

Figure~\ref{fig:torque-deltap_u} presents a comparison of non-dimensional torque and pressure-drop coefficients $T^*$ and $P^*$, as a function of $\phi$, for steady and periodic simulations. As expected, dynamic effects are negligible. Figure~\ref{fig:wells_entropy_u} presents a similar comparison for the non-dimensional entropy generation rate $K_{\dot{S}_G}$: again, in this case, hysteretic effects are minimal, and insufficient to generate any appreciable differences in the integral values of force coefficients.

\begin{figure}[!ht]
 \centering
 \includegraphics[trim={0 0 0 .0},clip,width=.75\columnwidth]{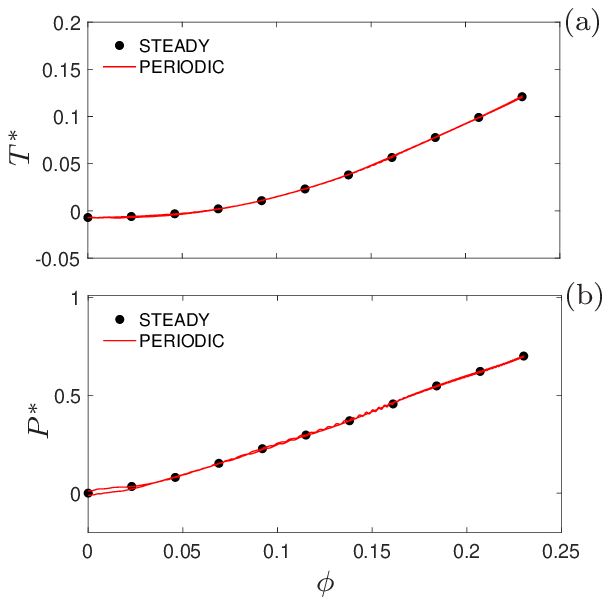}
\caption{Comparison of steady and periodic CFD results for non-dimensional torque $T^*$ (a) and static pressure drop $P^*$ (b) as a function of flow coefficient $\phi$}
\label{fig:torque-deltap_u}
\end{figure}

\begin{figure}[!ht]
 \centering
 \includegraphics[trim={0 0 0 .0},clip,width=.75\columnwidth]{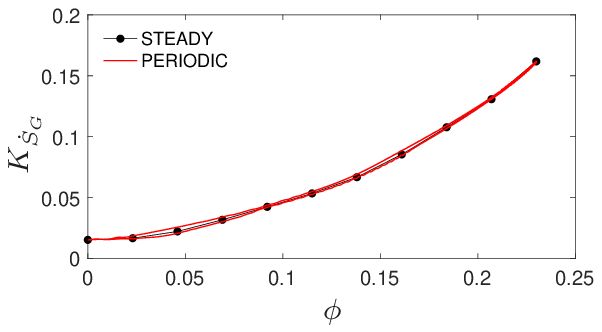}
\caption{Comparison of steady and periodic CFD results for the non-dimensional global entropy generation rate as a function of flow coefficient $\phi$}
\label{fig:wells_entropy_u}
\end{figure}
}

\section{Conclusions}
\label{sec:conclusion}

In recent years, a few authors have started to analyze entropy generation in Wells turbines, as a means to improve the overall performance of the system.
To the best of the authors' knowledge, all the applications of second law analyses to Wells turbines to date contain errors that lead to incorrect and unphysical results, such as an underestimation of the gradient of the entropy generation vs. velocity curve, or a reduction in losses for large incidence angles. These unphysical results are caused by an incorrect evaluation of the local entropy generation rate in Reynolds averaged conservation equations. 

This paper presents a critical analysis of the entropy generation mechanisms in fluid flow, with a focus on the effect of Reynolds averaging on conservation equations and on entropy generation in particular. The importance of turbulent fluctuations, previously neglected by other authors, is highlighted and a methodology for estimating their impact on entropy generation is presented. 

This theoretical analysis is then validated on two test cases: an isolated NACA0012 airfoil at different Reynolds numbers and incidence angles, and a Wells turbine. As for isolated airfoils a simple relationship between entropy generation and drag exists, this test case allows the approach to be verified before it is applied to a more complex problem. {\color{black} Most of the analyses are performed using steady simulations, given the low values of non-dimensional frequency Wells turbines operate at, which are not sufficient to produce appreciable dynamic effects. A comparison between a steady and an unsteady (periodic) analysis is presented in Section~\ref{sec:results_unsteady} to demonstrate the validity of this hypothesis.} 

The results highlight how in both cases the contribution of turbulent fluctuations is significant and cannot be neglected. Even more importantly, in the presence of regions of separated flow, the vast majority of entropy generation is due to turbulent dissipation, and considering only the contribution of the mean flow, as done in~\cite{Shehata2016,Soltanmohamadi2016}, would lead to the complete neglect of the main source of irreversibility, and hence decidedly incorrect results.

The use of second law analyses in CFD methods allows the identification of the main sources of
irreversibility in a fluid system. The paper demonstrates that this method, when correctly applied, can become an important basis for the analysis and optimization of the performance of Wells turbines and, in general, of turbomachinery.

\section*{Acknowledgements}

Funding: this work has been funded by the Regione Autonoma Sardegna under grant F72F16002880002 (L.R. 7/2007 n. 7 - year 2015).

Conflict of interest: the authors declare that they have no conflict of interest.
\nomenclature[O]{$\nabla$}{nabla operator [m$^{-1}$]}
\nomenclature[O]{$\nabla^2$}{laplacian operator [m$^{-2}$]}
\nomenclature[O]{$\nabla^s$}{sum of gradient and gradient transposed [m$^{-1}$]}
\nomenclature[O]{$\cdot$}{dot product}
\nomenclature[O]{$\otimes$}{cross product}
\nomenclature[O]{$\frac{D()}{Dt}$}{total time derivative [s$^{-1}$]}
\nomenclature[O]{$\frac{\partial()}{\partial t}$}{partial time derivative [s$^{-1}$]}
\nomenclature[O]{$\overline{()}$}{time average}
\nomenclature[O]{$\Delta$}{difference}
\nomenclature[Da]{${\bf{a}}$}{acceleration [m s$^{-2}$]}  
\nomenclature[Db]{$A$}{area [m$^{2}$]}
\nomenclature[Dc]{$c_p$}{constant pressure specific heat [m$^{2}$ s$^{-2}$ K$^{-1}$]}
\nomenclature[Dd]{$c_v$}{constant volume specific heat [m$^{2}$ s$^{-2}$ K$^{-1}$]}
\nomenclature[De]{$dS$}{differential surface [m$^{2}$]}
\nomenclature[Df]{$D$}{drag force [kg m s$^{-2}$]}
\nomenclature[Dg]{$f$}{frequency [s$^{-1}$]}
\nomenclature[Dh]{${\bf{F}}$}{force [kg m s$^{-2}$]}
\nomenclature[Di]{$h$}{specific enthalpy [m$^2$ s$^{-2}$]}   
\nomenclature[Dj]{$H$}{total specific enthalpy [m$^2$ s$^{-2}$]} 
\nomenclature[Dk]{$k$}{specific turbulent kinetic energy [m$^{2}$ s$^{-2}$]}
\nomenclature[Dm]{$p$}{pressure [kg m$^{-1}$ s$^{-2}$]}
\nomenclature[Dn]{${\bf{q}}$}{thermal flux [kg s$^{-3}$]}  
\nomenclature[Do]{$r_{tip}$}{turbine tip radius [m]}
\nomenclature[Dp]{$R$}{gas constant [m$^{2}$ s$^{-2}$ K$^{-1}$]}
\nomenclature[Dq]{$s$}{specific entropy [m$^{2}$ s$^{-2}$ K$^{-1}$]}
\nomenclature[Dr]{$S_1$}{near-field (surface) [m$^{2}$]} }
\nomenclature[Ds]{$S_2$}{far-field (surface) [m$^{2}$]}
\nomenclature[Dt]{$\dot{S}_G$}{entropy generation rate [kg m$^2$ s$^{-3}$ s$^{-1}$]}
\nomenclature[Du]{$t$}{time [s]}
\nomenclature[Dv]{$T$}{temperature [K], torque [kg m$^2$ s$^{-2}$]}
\nomenclature[Dz]{$u,{\mathbf{u}}$}{velocity [m s$^{-1}$]}
\nomenclature[Dza]{$V_a$}{turbine axial flow velocity [m s$^{-1}$]}
\nomenclature[Dzb]{$\epsilon$}{turbulent dissipation rate [m$^{2}$ s$^{-3}$]}
\nomenclature[Dzc]{$\Phi$}{dissipation function [s$^{-2}$]}
\nomenclature[Dzd]{$\lambda$}{thermal conductivity [kg m s$^{-3}$ K$^{-1}$]}   
\nomenclature[Dze]{$\lambda_t$}{turbulent conductivity [kg m s$^{-3}$ K$^{-1}$]}   
\nomenclature[Dzf]{$\mu$}{dynamic viscosity [kg m$^{-1}$ s$^{-1}$]}   
\nomenclature[Dzg]{$\mu_t$}{turbulent dynamic viscosity [kg m$^{-1}$ s$^{-1}$]}   
\nomenclature[Dzh]{$\omega$}{specific turbulent dissipation rate [s$^{-1}$], rotational speed [s$^{-1}$]} 
\nomenclature[Dzj]{$\Omega$}{mid-field (volume) [m$^{3}$]}
\nomenclature[Dzk]{${\bf{\Pi}}$}{viscous stress tensor [kg m$^{-1}$ s$^{-2}$]} 
\nomenclature[Dzl]{${\bf{\Pi}_R}$}{Reynolds stress tensor [kg m$^{-1}$ s$^{-2}$]} 
\nomenclature[Dzm]{$\rho$}{density [kg m$^{-3}$]}
\nomenclature[Dzn]{$\sigma_T$}{entropy generation rate per unit mass due to heat transfer [m$^{2}$ s$^{-3}$ K$^{-1}$]}
\nomenclature[Dzo]{$\sigma_V$}{entropy generation rate per unit mass due to fluid flow [m$^{2}$ s$^{-3}$ K$^{-1}$]}
\nomenclature[Na]{$c_d$}{drag coefficient}
\nomenclature[Nb]{$e$}{Nepero constant}
\nomenclature[Nc]{$k$}{non-dimensional frequency}
\nomenclature[Nd]{$K_{\dot{S}_G}$}{non-dimensional entropy generation rate}
\nomenclature[Ne]{$M$}{Mach number}
\nomenclature[Nf]{${\mathbf{n}}$}{surface normal unit vector}
\nomenclature[Ng]{$Re$}{Reynolds number}
\nomenclature[Nh]{$\alpha$}{incidence angle}
\nomenclature[Ni]{$\gamma$}{ratio of specific heats}
\nomenclature[Nj]{$\phi$}{turbine flow coefficient}
\nomenclature[Sa]{$'$}{fluctuating part}
\nomenclature[Sb]{$eff$}{effective}
\nomenclature[Sd]{$\infty$}{at infinity (upstream)}
\nomenclature[Sd]{$mf$}{mean flow}
\nomenclature[Sr]{$t$}{turbulent, due to turbulence}
\nomenclature[A]{CFD}{computational fluid dynamics}
\nomenclature[A]{RANS}{Reynolds Averaged Navier Stokes}
\nomenclature[A]{DNS}{direct numerical simulation}
\nomenclature[A]{LES}{large eddy simulation}

\begin{thenomenclature} 

 \nomgroup{A}

  \item [{CFD}]\tabto{1.3cm}\begingroup computational fluid dynamics\nomeqref {50}
		\nompageref{14}
  \item [{DNS}]\tabto{1.3cm}\begingroup direct numerical simulation\nomeqref {50}
		\nompageref{14}
  \item [{LES}]\tabto{1.3cm}\begingroup large eddy simulation\nomeqref {50}
		\nompageref{14}
  \item [{RANS}]\tabto{1.3cm}\begingroup Reynolds Averaged Navier Stokes\nomeqref {50}
		\nompageref{14}

 \nomgroup{D}

  \item [{${\bf{a}}$}]\tabto{1.3cm}\begingroup acceleration [m s$^{-2}$]\nomeqref {50}
		\nompageref{14}
  \item [{$A$}]\tabto{1.3cm}\begingroup area [m$^{2}$]\nomeqref {50}\nompageref{14}
  \item [{$c_p$}]\tabto{1.3cm}\begingroup constant pressure specific heat [m$^{2}$ s$^{-2}$ K$^{-1}$]\nomeqref {50}
		\nompageref{14}
  \item [{$c_v$}]\tabto{1.3cm}\begingroup constant volume specific heat [m$^{2}$ s$^{-2}$ K$^{-1}$]\nomeqref {50}
		\nompageref{14}
  \item [{$dS$}]\tabto{1.3cm}\begingroup differential surface [m$^{2}$]\nomeqref {50}
		\nompageref{14}
  \item [{$D$}]\tabto{1.3cm}\begingroup drag force [kg m s$^{-2}$]\nomeqref {50}
		\nompageref{14}
  \item [{$f$}]\tabto{1.3cm}\begingroup frequency [s$^{-1}$]\nomeqref {50}
		\nompageref{14}
  \item [{${\bf{F}}$}]\tabto{1.3cm}\begingroup force [kg m s$^{-2}$]\nomeqref {50}
		\nompageref{14}
  \item [{$h$}]\tabto{1.3cm}\begingroup specific enthalpy [m$^2$ s$^{-2}$]\nomeqref {50}
		\nompageref{14}
  \item [{$H$}]\tabto{1.3cm}\begingroup total specific enthalpy [m$^2$ s$^{-2}$]\nomeqref {50}
		\nompageref{14}
  \item [{$k$}]\tabto{1.3cm}\begingroup specific turbulent kinetic energy [m$^{2}$ s$^{-2}$]\nomeqref {50}
		\nompageref{14}
  \item [{$p$}]\tabto{1.3cm}\begingroup pressure [kg m$^{-1}$ s$^{-2}$]\nomeqref {50}
		\nompageref{14}
  \item [{${\bf{q}}$}]\tabto{1.3cm}\begingroup thermal flux [kg s$^{-3}$]\nomeqref {50}
		\nompageref{14}
  \item [{$r_{tip}$}]\tabto{1.3cm}\begingroup turbine tip radius [m]\nomeqref {50}
		\nompageref{14}
  \item [{$R$}]\tabto{1.3cm}\begingroup gas constant [m$^{2}$ s$^{-2}$ K$^{-1}$]\nomeqref {50}
		\nompageref{14}
  \item [{$s$}]\tabto{1.3cm}\begingroup specific entropy [m$^{2}$ s$^{-2}$ K$^{-1}$]\nomeqref {50}
		\nompageref{14}
  \item [{$S_1$}]\tabto{1.3cm}\begingroup near-field (surface) [m$^{2}$]\nomeqref {50}
		\nompageref{14}
  \item [{$S_2$}]\tabto{1.3cm}\begingroup far-field (surface) [m$^{2}$]\nomeqref {50}
		\nompageref{14}
  \item [{$\dot{S}_G$}]\tabto{1.3cm}\begingroup entropy generation rate [kg m$^2$ s$^{-3}$ s$^{-1}$]\nomeqref {50}
		\nompageref{14}
  \item [{$t$}]\tabto{1.3cm}\begingroup time [s]\nomeqref {50}\nompageref{14}
  \item [{$T$}]\tabto{1.3cm}\begingroup temperature [K], torque [kg m$^2$ s$^{-2}$]\nomeqref {50}
		\nompageref{14}
  \item [{$u,{\mathbf{u}}$}]\tabto{1.3cm}\begingroup velocity [m s$^{-1}$]\nomeqref {50}
		\nompageref{14}
  \item [{$V_a$}]\tabto{1.3cm}\begingroup turbine axial flow velocity [m s$^{-1}$]\nomeqref {50}
		\nompageref{14}
  \item [{$\epsilon$}]\tabto{1.3cm}\begingroup turbulent dissipation rate [m$^{2}$ s$^{-3}$]\nomeqref {50}
		\nompageref{14}
  \item [{$\Phi$}]\tabto{1.3cm}\begingroup dissipation function [s$^{-2}$]\nomeqref {50}
		\nompageref{14}
  \item [{$\lambda$}]\tabto{1.3cm}\begingroup thermal conductivity [kg m s$^{-3}$ K$^{-1}$]\nomeqref {50}
		\nompageref{14}
  \item [{$\lambda_t$}]\tabto{1.3cm}\begingroup turbulent conductivity [kg m s$^{-3}$ K$^{-1}$]\nomeqref {50}
		\nompageref{14}
  \item [{$\mu$}]\tabto{1.3cm}\begingroup dynamic viscosity [kg m$^{-1}$ s$^{-1}$]\nomeqref {50}
		\nompageref{14}
  \item [{$\mu_t$}]\tabto{1.3cm}\begingroup turbulent dynamic viscosity [kg m$^{-1}$ s$^{-1}$]\nomeqref {50}
		\nompageref{14}
  \item [{$\omega$}]\tabto{1.3cm}\begingroup specific turbulent dissipation rate [s$^{-1}$], rotational speed [s$^{-1}$]\nomeqref {50}
		\nompageref{14}
  \item [{$\Omega$}]\tabto{1.3cm}\begingroup mid-field (volume) [m$^{3}$]\nomeqref {50}
		\nompageref{14}
  \item [{${\bf{\Pi}}$}]\tabto{1.3cm}\begingroup viscous stress tensor [kg m$^{-1}$ s$^{-2}$]\nomeqref {50}
		\nompageref{14}
  \item [{${\bf{\Pi}_R}$}]\tabto{1.3cm}\begingroup Reynolds stress tensor [kg m$^{-1}$ s$^{-2}$]\nomeqref {50}
		\nompageref{14}
  \item [{$\rho$}]\tabto{1.3cm}\begingroup density [kg m$^{-3}$]\nomeqref {50}
		\nompageref{14}
  \item [{$\sigma_T$}]\tabto{1.3cm}\begingroup entropy generation rate per unit mass due to heat transfer [m$^{2}$ s$^{-3}$ K$^{-1}$]\nomeqref {50}
		\nompageref{14}
  \item [{$\sigma_V$}]\tabto{1.3cm}\begingroup entropy generation rate per unit mass due to fluid flow [m$^{2}$ s$^{-3}$ K$^{-1}$]\nomeqref {50}
		\nompageref{14}

 \nomgroup{N}

  \item [{$c_d$}]\tabto{1.3cm}\begingroup drag coefficient\nomeqref {50}
		\nompageref{14}
  \item [{$e$}]\tabto{1.3cm}\begingroup Nepero constant\nomeqref {50}\nompageref{14}
  \item [{$k$}]\tabto{1.3cm}\begingroup non-dimensional frequency\nomeqref {50}
		\nompageref{14}
  \item [{$K_{\dot{S}_G}$}]\tabto{1.3cm}\begingroup non-dimensional entropy generation rate\nomeqref {50}
		\nompageref{14}
  \item [{$M$}]\tabto{1.3cm}\begingroup Mach number\nomeqref {50}\nompageref{14}
  \item [{${\mathbf{n}}$}]\tabto{1.3cm}\begingroup surface normal unit vector\nomeqref {50}
		\nompageref{14}
  \item [{$Re$}]\tabto{1.3cm}\begingroup Reynolds number\nomeqref {50}
		\nompageref{14}
  \item [{$\alpha$}]\tabto{1.3cm}\begingroup incidence angle\nomeqref {50}
		\nompageref{14}
  \item [{$\gamma$}]\tabto{1.3cm}\begingroup ratio of specific heats\nomeqref {50}
		\nompageref{14}
  \item [{$\phi$}]\tabto{1.3cm}\begingroup turbine flow coefficient\nomeqref {50}
		\nompageref{14}

 \nomgroup{O}

  \item [{$\cdot$}]\tabto{1.3cm}\begingroup dot product\nomeqref {50}\nompageref{14}
  \item [{$\Delta$}]\tabto{1.3cm}\begingroup difference\nomeqref {50}\nompageref{14}
  \item [{$\frac{\partial()}{\partial t}$}]\tabto{1.3cm}\begingroup partial time derivative [s$^{-1}$]\nomeqref {50}
		\nompageref{14}
  \item [{$\frac{D()}{Dt}$}]\tabto{1.3cm}\begingroup total time derivative [s$^{-1}$]\nomeqref {50}
		\nompageref{14}
  \item [{$\nabla$}]\tabto{1.3cm}\begingroup nabla operator [m$^{-1}$]\nomeqref {50}
		\nompageref{14}
  \item [{$\nabla^2$}]\tabto{1.3cm}\begingroup laplacian operator [m$^{-2}$]\nomeqref {50}
		\nompageref{14}
  \item [{$\nabla^s$}]\tabto{1.3cm}\begingroup sum of gradient and gradient transposed [m$^{-1}$]\nomeqref {50}
		\nompageref{14}
  \item [{$\otimes$}]\tabto{1.3cm}\begingroup cross product\nomeqref {50}
		\nompageref{14}
  \item [{$\overline{()}$}]\tabto{1.3cm}\begingroup time average\nomeqref {50}
		\nompageref{14}

 \nomgroup{S}

  \item [{$'$}]\tabto{1.3cm}\begingroup fluctuating part\nomeqref {50}
		\nompageref{14}
  \item [{$eff$}]\tabto{1.3cm}\begingroup effective\nomeqref {50}\nompageref{14}
  \item [{$\infty$}]\tabto{1.3cm}\begingroup at infinity (upstream)\nomeqref {50}
		\nompageref{14}
  \item [{$mf$}]\tabto{1.3cm}\begingroup mean flow\nomeqref {50}\nompageref{14}
  \item [{$t$}]\tabto{1.3cm}\begingroup turbulent, due to turbulence\nomeqref {50}
		\nompageref{14}

\end{thenomenclature}
\bibliographystyle{spmpsci}

\end{document}